\documentclass[aps,pra,twocolumn,groupedaddress,nofootinbib,notitlepage,showpacs,floatfix]{revtex4}

\usepackage{graphicx,graphics,epsfig,times,bm,bbm,amssymb,amsmath,amsfonts,mathrsfs,color}
\usepackage{hyperref}
\usepackage[normalem]{ulem}
\usepackage{float}
\usepackage{ifpdf}

\restylefloat{figure,captions}

\newcommand{\ig}[1]{}
\def\s0{I}
\def\sx{\sigma^x}

\def\sz{\sigma^z}
\newcommand{\red}[1]{\textcolor{black}{#1}} 
\newcommand{\green}[1]{\textcolor{green}{#1}} 
\newcommand{\beq} {\begin{equation}}
\newcommand{\eeq} {\end{equation}}
\newcommand{\bea} {\begin{eqnarray}}
\newcommand{\eea} {\end{eqnarray}}

\begin{document}

\title{Quadratic Dynamical Decoupling with Non-Uniform Error Suppression}
\author{Gregory Quiroz}
\affiliation{Department of Physics and Center for Quantum Information Science \&
Technology, University of Southern California, Los Angeles, California
90089, USA}
\author{Daniel A. Lidar}
\affiliation{Departments of Electrical Engineering, Chemistry, and Physics, and Center
for Quantum Information Science \& Technology, University of Southern
California, Los Angeles, California 90089, USA}

\begin{abstract}
We analyze numerically the performance of the near-optimal quadratic
dynamical decoupling (QDD) single-qubit decoherence errors suppression
method [J. West \textit{et al}., Phys. Rev. Lett. \textbf{104}, 130501
(2010)]. The QDD sequence is formed by nesting two optimal Uhrig dynamical
decoupling sequences for two orthogonal axes, comprising $N_1$ and $N_2$
pulses, respectively. Varying these numbers, we study the decoherence
suppression properties of QDD directly by isolating the errors associated
with each system basis operator present in the system-bath interaction
Hamiltonian. Each individual error scales with the lowest order of the Dyson
series, therefore immediately yielding the order of decoherence suppression.
We show that the error suppression properties of QDD are dependent upon the
parities of $N_1$ and $N_2$, and near-optimal performance is achieved for
general single-qubit interactions when $N_1=N_2$.
\end{abstract}

\pacs{03.67.Pp, 03.65.Yz, 82.56.Jn, 76.60.Lz}

\maketitle

\section{Introduction}

In recent years, there have been promising advances towards the usage of
quantum systems to perform quantum information processing (QIP) \cite%
{Ladd:10}. However, for these systems to be utilized efficiently, it is
necessary to preserve and store information for a sufficient amount of time
so that computations can be implemented. Unfortunately, quantum systems are
generally hindered in their ability to perform such tasks effectively due to
unwanted interactions between the system and its enviroment, which result in
decoherence \cite{Schlosshauer:book}.

Dynamical Decoupling (DD) is a strategy that can be used to suppress
decoherence and effectively remove unwanted system-environment interactions
for a period of time such that quantum states can be preserved with a low
probability of error \cite{Viola:99}. As originally conceived, DD schemes
are characterized by the application of short control pulses to the system
such that the overall time evolution provided by the pulses selectively
average out system-environment interactions, thereby suppressing decoherence 
\cite{Viola:98,Duan:98e,Ban:98,Zanardi:98b}. One advantage of DD\ is that it
is an open-loop control method, which does not require any measurements or
feedback, unlike quantum error correction \cite{Gaitan:book}. Nor does DD
require any specific knowledge of the environment other than it being
non-Markovian, unlike optimal control methods designed to suppress
decoherence \cite{GKL:08,Clausen:10}.

Early DD\ schemes were designed to remove unwanted system-bath interactions
to a given, low order in time-dependent perturbation theory \cite{Viola:99}.
Concatenated DD (CDD) was the first explicit scheme capable of removing such
interactions to an arbitrary order \cite{KhodjastehLidar:04}. CDD
accomplishes this via a recursive construction in which each successive
level removes another order in time-dependent perturbation theory. The
advantages of CDD over standard periodic pulse sequences have been
extensively studied analytically \cite{KhodjastehLidar:07,NLP:09} and
numerically \cite{KhodjastehLidar:07,Witzel:07a,PhysRevB.75.201302,Zhang:08,West:10}, and
confirmed in a number of recent experimental studies \cite{Peng:11,Alvarez:10,2010arXiv1011.1903T,2010arXiv1011.6417W,2010arXiv1007.4255B}. However, assuming that pulse intervals can be made arbitrarily short, the
number of pulses required to achieve arbitrary order suppression grows
exponentially with the order in CDD. When the finiteness of pulse intervals
is accounted for there is an optimal level of concatenation and
correspondingly a highest attainable perturbation theory order for removal
of unwanted interactions \cite{KhodjastehLidar:07,NLP:09,Alvarez:10}.

In contrast to CDD, Uhrig DD (UDD) is characterized by the use of unequal
pulse intervals, or free evolution periods \cite{Uhrig:07}. By applying control pulses at 
\begin{equation}
t_j=T \sin^2\frac{j\pi}{2(N+1)},  \label{eq:UDDIntervals}
\end{equation}
where $j=1,2,\ldots,N+1$, a UDD sequence of total duration $T$ yields
$N$th order decoupling for single-axis
system-bath coupling \cite{yang:180403}. The number of pulses comprising a UDD sequence is $N$ if $N$ is even, or $N+1$ if $N$ is odd. Generalizations of UDD for generic
system-environment interactions include Concatenated UDD (CUDD) \cite{CUDD},
Quadratic Dynamical Decoupling (QDD) \cite{WFL:09} and Nested Uhrig
Dynamical Decoupling (NUDD) \cite{NUDD} (see also \cite{Mukhtar:10}).
CUDD removes the restriction of single-axis decoupling and suppresses
general (three-axis) decoherence errors on a qubit, but still suffers from
the exponential cost of CDD. On the other hand, QDD also suppresses general
decoherence errors on a qubit, but does so without the exponential cost of CDD
by nesting two UDD sequences for two orthogonal axes. In fact, QDD is a
near-optimal scheme for single-qubit decoupling from an arbitrary bath,
requiring only $N^2$ pulses for $N$th order decoupling if $N$ is even, or  $(N+1)^2$ pulses if $N$ is odd. The NUDD sequence is
built on the same nesting idea, but removes the QDD restriction of
single-qubit decoupling. NUDD applies to arbitrary multi-level systems
coupled to arbitrary baths, as recently proved in Ref.~\cite{Jiang:11}.

In this work we focus on the decoherence errors suppression capabilities of
QDD. This protocol requires two nested sequences each containing $N_j$ pulses if $N_j$ is even or $N_j+1$
pulses if $N_j$ is odd, $j=1,2$. We call $N_1$ and $N_2$ the inner and outer sequence order,
respectively. While the original QDD paper \cite{WFL:09} noted that
sequences with $N_1 \neq N_2$ are possible and could be advantageous when a
particular axis is dominant, only the case $N_1=N_2$ was analyzed. Here we
numerically study QDD for $N_1\neq N_2$. We provide a complete numerical
elucidation of the performance of QDD as a function of $N_1$ and $N_2$.

The structure of this paper is as follows. In Section~\ref{synopsis} we
provide a brief synopsis of our results. In Section~\ref{protocol} we
summarize the QDD protocol. In Section~\ref{QDDperformance} we introduce the relevant error measures utilized in
this paper and provide an analysis of the expected scaling of these measures with the inner and outer sequence orders. Section ~\ref{Numerics} is devoted to our numerical results. In it we discuss the scaling of the single-axis errors and their time-dependence, as well as the scaling of a distance measure for the entire QDD sequence. \ig{and }Section~\ref{Conclusion} presents our conclusions.

\section{Synopsis of results}
\label{synopsis}

We characterize QDD performance with respect to the order of error
suppression given by the overall fidelity loss of the qubit (system) state.
We show that the order of overall error suppression is dictated by the
lowest of the inner or outer sequence orders, namely $D \sim
(J\tau)^{\min\{N_1,N_2\}+1}$, where $D$ is a measure of the overall error, $%
J $ is the strength of the system-bath coupling, and $\tau$ is the smallest
pulse interval.

To arrive at these results and gain more insight we \ig{also }first  characterize QDD performance with respect to
the order of error suppression given by the order of error suppression along
each axis of the qubit Bloch sphere, referred to as the single-axis error.
We isolate the single-axis errors by projecting the total evolution operator
into the three directions defined by the Pauli basis. The error suppression properties are then
distinguished with respect to the scaling of the error as a function of the
minimum pulse interval for various inner and outer sequence orders. Since
this scaling is dominated by the first non-zero term of the Dyson series,
the order of error suppression for each single-axis error can be
characterized with respect to inner and outer sequence order.  Suppression of
the $x$-axis error (or ``$X$-type error") is determined by $N_1$, the $y$-axis 
error by the parity of $N_1$ and $N_2$, and the $z$-axis error by $N_2$ except
in the case when $N_1$ is odd [see Eqs.~\eqref{eq:nx}-\eqref{eq:ny}]. Our results will show that if $N_1$ is odd, there
is a constraint on the suppression of the $z$-axis error depending on the value of
$N_1$ with respect to $N_2$.

Parity effects were anticipated in Ref.~\cite{NUDD}, where the expected
performance of QDD was proved for sequences with even $N_1$ (the proof was recently completed in Ref.~\cite{Jiang:11}). We show that parity effects are absent in QDD
\ig{displays a dependence on the parity of the sequence orders only for
interactions anti-commuting with both inner and outer sequence decoupling
operators. Interactions} only for the interaction that anti-commutes solely with the decoupling operator
comprising the inner sequence. This interaction ($\sx$) is \ig{are shown to be} suppressed with UDD
efficiency, i.e., $N_1$th order error suppression for a pulse sequence
comprising $N_1$ pulses. This result holds for the outer sequence as well (i.e., the $\sz$ interaction,
except in the case of $N_2\geq 2(N_1+1)$ with $N_1$ odd, where the inner sequence hinders the ability
of the outer sequence to suppress decoherence to the expected order. We find that in general, UDD
efficiency for general single-qubit errors is achieved when $N_1$ and $N_2$
are both even.

We introduce another perspective on the QDD sequence, by studying the time-dependence of the single-axis errors. We find that as the sequence progresses, these errors oscillate between values that are near to their final minimum, and much higher values. \ig{An interesting aspect of this result is that if one is interested in suppressing just one particular type of single-axis error, there is no need to wait until the conclusion of the entire sequence.}

\section{QDD Protocol}

\label{protocol}

UDD suppresses single-qubit dephasing or longitudinal relaxation errors 
\emph{separately}, for a general environment \cite{Yang:08,UL:10}. The
extension of UDD to QDD improves on this by handling general single-qubit
decoherence, in particular both dephasing and relaxation simultaneously.
Therefore, in our analysis of QDD the time-independent Hamiltonian 
\begin{eqnarray}
H &=&H_{B}+H_{SB},  \label{eq:H} \\
H_{B} &=&\s0\otimes B_{I},  \label{eq:HB} \\
H_{SB} &=&\sigma ^{x}\otimes B_{x}+\sigma ^{y}\otimes B_{y}+\sigma
^{z}\otimes B_{z}  \label{eq:HSB}
\end{eqnarray}%
is employed to describe general system-environment interactions for the
single-qubit system. The system operators $\sigma^{\mu }$ are the standard
Pauli matrices, while the bounded operators, $B_{\mu }$, $\mu \in
\{I,x,y,z\} $, characterize a generic environment. The operator $B_{I}$
encompasses the pure bath dynamics, so that $H_{B}$ is the \textquotedblleft
pure-bath\textquotedblright\ Hamiltonian, while $H_{SB}$ is the system-bath
interaction Hamiltonian.

In building a QDD sequence to compensate for the interactions present in the
above Hamiltonian, it is useful to first choose a so-called Mutually
Orthogonal Operator Set (MOOS) \cite{NUDD}. This set comprises mutually
anti-commuting or commuting operators needed to perform DD in the general
NUDD scheme. Each of these operators is required to anti-commute with some
portion of the interaction Hamiltonian. The anti-commutation condition
between an element of the MOOS and the interaction Hamiltonian is essential
for decoherence suppression. One might expect a UDD sequence composed of an element of the
MOOS \ig{is expected} to suppress the corresponding anti-commuting interaction
with UDD efficiency. As we shall show, this is not the case, except for the inner sequence.

In the case of a single-qubit system, NUDD reduces to QDD and the MOOS
requires only two operators. There is some flexibility in choosing the MOOS
for Eq.~(\ref{eq:H}), but without loss of generality we pick $\Omega =\{X,Z\}$\ig{ will
be the specified MOOS here}. The operators $X=\sigma ^{x}\otimes I$ and $Z=\sigma ^{z}\otimes I$ (we drop global phase factors everywhere) represent ideal zero-width $\pi $-rotations about their respective
axes of the qubit subspace, and do not affect the bath\ig{, by assumption}. The QDD sequence is now readily constructed as 
\cite{WFL:09} 
\begin{equation}
U_{\Omega }^{(N_{1},N_{2})}=X^{N_{2}}Z^{(N_{1})}(s_{N_{2}+1}\tau )\cdots
XZ^{(N_{1})}(s_{1}\tau ),  
\label{eq:QDD}
\end{equation}%
where 
\begin{equation}
Z^{(N_{1})}(\tau )=Z^{N_{1}}U(s_{N_{1}+1}\tau )\cdots ZU(s_{2}\tau 
)ZU(s_{1}\tau ).  
\label{eq:Zseq}
\end{equation}%
The free evolution dynamics
between successive pulses, $U(t)=e^{-iHt}$, is governed by Eq.~(\ref{eq:H}),
and the free evolution time durations are given in terms of the normalized
UDD intervals, 
\begin{equation}
s_{j}=\frac{t_{j}-t_{j-1}}{t_{1}-t_{0}},
\end{equation}%
with $t_{j}$ specified by Eq.~(\ref{eq:UDDIntervals}), and $\tau = t_1-t_0$.  The high efficiency of UDD in suppressing decoherence, and therefore QDD,
arises from the choice of the relative free evolution time durations $\{s_j\}$.

The total normalized time
of an $N$-pulse UDD sequence is given by,
\begin{equation}
\label{Tn}
  S^{(N)} \equiv \sum_{j=1}^{N+1} s_j = \frac{t_{N+1}}{t_1} =
  \csc^2\left(\frac{\pi}{2N+2}\right),
\end{equation}
so that the total physical time is 

\begin{equation}
T^{(N)} = S^{(N)} \tau.
\end{equation}

Therefore the total normalized time
of a QDD sequence with $N_1$ inner and $N_2$ outer pulses is given by,
\begin{eqnarray}
\label{Tn1n2}
  S^{(N_1,N_2)} &\equiv& \sum_{j=1}^{N_2+1} s_j S^{(N_1)} \notag \\
  &=& 
  \csc^2\left(\frac{\pi}{2N_1+2}\right) \csc^2\left(\frac{\pi}{2N_2+2}\right),
\end{eqnarray}
so that the total physical time is $T^{(N_1,N_2)} = S^{(N_1,N_2)} \tau$.

\section{QDD performance measures}
\label{QDDperformance}
\subsection{QDD analysis}

\ig{[Greg, note ignored text in the tex file here]}
\ig{As mentioned above, it has been proven that UDD has the ability to suppress
general 
single-axis system-bath interactions with high efficiency \cite{Yang:08,UL:10}. However no analogous proof is currently known for
QDD (a partial proof was given in Ref.~\cite{NUDD}).
While such a proof is forthcoming \cite{KL:11}.}
Our goal here is to
understand the properties of QDD error suppression for general $N_{1}$ and $N_{2}$ and for a wide range of parameters. We do so by isolating the errors proportional to each system basis
operator, $\sigma ^{\mu }$, $\mu =\{x,y,z\}$, i.e., the single-axis errors.
In this manner the order of error suppression can be extracted directly and
possible constraints on QDD effectiveness can be accurately identified. Each
single-axis error is obtained from the evolution operator, $U_{\Omega
}^{(N_{2},N_{1})}$, by projecting along the particular axis of interest and
performing a partial trace over the system. The order of error suppression
can then be quantified by the scaling of the single-axis error as a function of
either the total evolution time or the minimum pulse interval. We choose the
minimum pulse interval since experimentally this quantity is always
lower-bounded, and plays an important role in the ultimate performance
limits of UDD \cite{UL:10} and DD in general \cite{PhysRevA.83.020305}.

The resulting QDD evolution operator, $U_{\Omega }^{(N_{2},N_{1})}$,
contains all the information regarding decoherence suppression for each
single-axis error. Construction of the final evolution operator is
accomplished by first considering the inner sequence evolution, $%
Z^{(N_{1})}(\tau )$. Let Eq.~(\ref{eq:H}) be partitioned such that $%
H=H_{+}+H_{-}$, where 
\begin{equation}
H_{+}=\sigma ^{x}\otimes B_{x}+\sigma ^{y}\otimes B_{y}
\end{equation}%
and 
\begin{equation}
H_{-}=\s0\otimes B_{I}+\sigma ^{z}\otimes B_{z}.
\end{equation}%
Clearly, the element of the MOOS comprising $Z^{(N_{1})}(\tau )$
anti-commutes with $H_{+}$ and commutes with $H_{-}$, i.e., $[H_{\pm
},Z]_{\pm }=0$, where the plus and minus sign subscripts signify the
anti-commutator and commutator, respectively.

The $Z$-type UDD sequence $Z^{(N_{1})}(\tau )$ is effective against the
anti-commuting Hamiltonian, $H_{+}$. $Z^{(N_{1})}(\tau )$ \ig{will be} is completely
ineffective against \ig{any} unwanted interactions within $H_{-}$. Any additional
errors associated with $H_{-}$ \ig{will have to} must be addressed using another
member of the MOOS. The inner sequence evolution can be expanded in terms of 
$H_{\pm }$ \cite{yang:180403}, 
\begin{eqnarray}
Z^{(N_{1})}(\tau ) &=&e^{-i[(-1)^{N_{1}}H_{+}+H_{-}]s_{N_{1}+1}\tau }\cdots 
\notag \\
&&\cdots e^{-i[-H_{+}+H_{-}]s_{2}\tau }e^{-i[H_{+}+H_{-}]s_{1}\tau },
\label{eq:Z}
\end{eqnarray}%
where the anti-commuting and commuting properties of $H_{\pm }$,
respectively, have been used. Transforming into the interaction picture with
respect to $H_{-}$, we can write $Z^{(N_{1})}(\tau )=U^{(N_{1})}_{-}(\tau )U_{z}^{(N_{1})}(\tau) $, such that $U_{-}^{(N_{1})}(\tau )=e^{-iH_{-}S^{(N_1)}\tau }$ and 
\begin{equation}
U_{z}^{(N_{1})}(\tau )=\hat{\mathcal{T}}\text{exp}\left( -i\int_{0}^{S^{(N_1)}\tau
}f_{z}(t)H_{+}(t)dt\right) .
\end{equation}
The modulation function { $f_{z}(t)=(-1)^{j-1}$} is defined for {$t\in \lbrack
\sum_{\ell=1}^{j-1}s_{\ell}\tau ,\sum_{\ell=1}^{j} s_{\ell}\tau ]$} and $\hat{\mathcal{T}}$ is the time-ordering
operator. $H_{+}$ in the rotating frame with respect to $H_{-}$ takes on the
form of a power series expansion in $t$, 
\begin{equation}
H_{+}(t)=U_{-}^{{(N_{1})}\dagger }(t)H_{+}U_{-}^{(N_{1})}(t)=\sum_{k=0}^{\infty
}H_{+}^{(k)}t^{k}.
\end{equation}%
The power series form of $H_{+}(t)$ is useful (though not essential \cite%
{NUDD}) for the proof of UDD and therefore the suppression of error
associated with $H_{+}$ \cite{Yang:08,UL:10,Pasini:09,2011arXiv1101.5286W}.
All constants of the expansion are condensed within $H_{+}^{(k)}$, along
with the $k$-fold commutator 
\begin{equation}
\lbrack _{k}H_{-},H_{+}]=[H_{-},[H_{-},\cdots \lbrack H_{-},H_{+}]\cdots ]].
\end{equation}%
Using time-dependent perturbation theory, $U_{z}^{(N_{1})}(\tau )$ is expanded in the
Dyson series 
\begin{equation}
U_{z}^{(N_{1})}(\tau )=\sum_{n=0}^{\infty }\sum_{\mathbf{k}_{n}}H_{+}^{(k_{n})}\cdots
H_{+}^{(k_{1})}F_{z}^{(\mathbf{k}_{n})}(\tau ),
\end{equation}%
where $\mathbf{k}_{n}=\{k_{1},...,k_{n}\}$ with $k_{i}=0,1,...$ for all $i$,
and all of the time-dependence of the expansion has been placed in 
\begin{equation}
F_{z}^{({N_{1},}\mathbf{k}_{n})}(\tau )=(-i)^{n}\int_{0}^{S^{(N_1)}\tau
}\int_{0}^{t_{n-1}}\cdots
\int_{0}^{t_{2}}\prod_{j=1}^{n}dt_{j}\,f_{z}(t_{j})t_{j}^{k_{j}}
\end{equation}%
The proof of UDD is completed by parametrizing $t_{j}$ as $t_{j}=\tau \sin
^{2}(\theta _{j}/2)$, Fourier expanding $f_{z}(t_{j})$, and showing that $%
F_{z}^{(N_1,\mathbf{k}_{n})}(\tau )=0$ for all odd values of $n$ {when $n+\sum^{n}_{j=1}k_j\leq N_1$} \cite{yang:180403}. All even orders of the expansion are
proportional to unity or $\sigma ^{z}$, and are therefore not associated
with $H_{+}$. The expansion ultimately yields 
\begin{equation}
Z^{(N_{1})}(\tau )=e^{-iH_{-}^{\prime }(\tau )S^{(N_1)}\tau +\mathcal{O}\left( (\tau
\Vert H_{+}^{\prime }(\tau )\Vert )^{N_{1}+1}\right) },  \label{eq:Uz}
\end{equation}%
where ${H}_{+}^{\prime }(\tau )$ is a generic single-qubit system-bath
Hamiltonian and 
\begin{equation}
H_{-}^{\prime }(\tau )=B_{I}^{\prime }(\tau )+\sigma ^{z}\otimes
B_{z}^{\prime }(\tau )
\end{equation}%
is composed of environment operators dependent on the minimum pulse
interval. Note that these environment operators are not the same operators
defined in Eq.~(\ref{eq:H}), but combinations of the original Hamiltonian
operators resulting from the perturbation expansion.

Left with only dephasing errors, terms proportional to $\sigma ^{z}$, the
process is continued again by defining $\tilde{H}_{+}=\sigma ^{z}\otimes
B_{z}^{\prime }(\tau )$ and $\tilde{H}_{-}=B_{I}^{\prime }(\tau )$ for the
outer $X$-type UDD sequence of Eq.~(\ref{eq:QDD}). $\tilde{H}_{\pm }$ is
defined in this way such that $[\tilde{H}_{\pm },X]_{\pm }=0$, a similar
condition to that required for the inner $Z$-type sequence. The resulting
evolution of Eq. (\ref{eq:QDD}) can be summed up as 
\begin{equation}
U_{\Omega }^{(N_{1},N_{2})}=B_{I}^{\prime \prime }(\tau )+\sum_{\mu \in
\{x,y,z\}}\sigma ^{\mu }\otimes B_{\mu }^{\prime \prime }(\tau ).
\label{eq:Uf}
\end{equation}
Once again, the time-dependent environment operators, $B^{\prime \prime
}_{\mu}(\tau)$, differ from the previously defined environment operators.
Each $B^{\prime \prime }_{\mu}(\tau)$, $\mu\in\{x,y,z\}$, contains the
uncompensated decoherence along each of the qubit axes. Bounds on the order of error
suppression are derived analytically in Ref.~\cite{KL:11}.

\subsection{Single-axis errors}

One of our goals is to characterize the performance of QDD with respect to
the remaining system-environment interaction operators, $B^{\prime \prime
}_{\nu}(\tau)$. The error is quantified by what we refer to as the
single-axis error $E_\mu$: 
\begin{equation}
E_{\mu}(\tau)=\|B^{{\prime \prime }}_{\mu}(\tau)\|_F,  
\label{E_mu}
\end{equation}
where 
\begin{equation}
B^{{\prime \prime }}_{\mu}(\tau)=\text{Tr}_{S}\left(U^{(N_1,N_2)}_{\Omega}%
\sigma^{\mu}\right),
\end{equation}
and where $\|A\|_F$ is the Frobenius norm of $A$, i.e., 
\begin{equation}
\|A\|_F = \mathrm{Tr}\sqrt{A^\dagger A},
\end{equation}
the sum of singular values of $A$. (The choice of norm is somewhat
arbitrary; we could have used any other unitarily invariant norm \cite%
{Bhatia:book}.) Thus $E_{\mu}(\tau)\geq 0$, and we are interested in how the
single-axis errors scale\ig{s with} as a function of the minimum pulse interval.

\subsection{Distance measure}

The advantage of the single-axis error over other measures such as
polarization or fidelity is that these typically scale with the overall
minimum order of decoherence suppression and therefore do not provide
detailed information about the structure of the unitary evolution operator
itself. However, the single-axis errors of course do not tell the whole
story of QDD performance. A useful overall fidelity-loss measure is the
distance \cite{Grace:10} 
\begin{equation}
D(U,G)=\frac{1}{\sqrt{d_S d_B}}\,\,\min_{\Phi}\|U-G\otimes\Phi\|_{F},
\label{eq:Dist}
\end{equation}
Here $d_S$ and $d_B$ denote the dimensions of the system and bath Hilbert
spaces, respectively, $U$ is the actual system-bath unitary evolution
operator, \ig{and} $G$ is the desired system-only unitary operator, and $\Phi$ is a bath operator. Grace {\it et al}.  \cite{Grace:10} give an explicit form for this distance measure, so that in our numerical simulations we do not need to compute the minimum over $\Phi$. In our case $%
G=I$ is the desired system unitary operator since the goal of DD is
to remove the system-environment interaction while effectively acting
trivially on the system. The advantage of the distance measure $D(U,G)$ of
Eq.~(\ref{eq:Dist}) over the standard Uhlman fidelity or trace-norm distance 
\cite{Nielsen:book} is that it is state-independent and can be correlated
directly with the results obtained for the single-axis errors.

We shall show that $D(U,I)$ scales in the same manner as $\min_\mu E_{\mu}(\tau) $. It will also illuminate some interesting features of QDD
based on the inner sequence order not captured by the single-axis errors.

\subsection{Scaling}

We parametrize the strength of the pure environment dynamics and
system-environment interactions, respectively, as 
\begin{equation}
\beta =\Vert H_{B}\Vert ,\quad J=\Vert H_{SB}\Vert , \quad J_\alpha =
\|B_\alpha \|  \label{eq:J}
\end{equation}%
where $\Vert A\Vert $ is the standard sup-operator norm, namely the largest
singular value (largest eigenvalue of $\sqrt{A^{\dagger }A}$):%
\begin{equation}
\Vert A\Vert =\sup_{|\psi \rangle }\frac{\sqrt{\langle \psi |A^{\dagger
}A|\psi \rangle }}{\sqrt{\langle \psi |\psi \rangle }}.
\end{equation}%
The effectiveness of DD tends to be greater in the regime where the
environment is essentially static and the duration of the free evolution is
much smaller than the environment correlation time: $J\tau \ll 1$ and $\beta
\ll J$.

We model the environment as a four-qubit bath with operators 
\begin{equation}
B_{\mu }=\sum_{i\neq j}\sum_{\alpha ,\beta }c_{\alpha \beta }^{\mu }\left(
\sigma _{i}^{\alpha }\otimes \sigma _{j}^{\beta }\right)   \label{eq:B_mu}
\end{equation}%
characterizing the dynamics of the bath and system-bath interactions. The
operators $B_{\mu }$ are composed of one- and two-body terms, where $i,j$
index the bath qubits, $\mu ,\alpha ,\beta \in \{1,x,y,\green{z}\}$, where $\sigma
^{1}=I$ is the $2\times 2$ identity matrix, and $c_{\alpha \beta }^{\mu }\in
\lbrack 0,1]$ are coefficients chosen uniformly at random. Constructing $%
B_{\mu }$ in this manner permits general two- and three-body interactions
between the system and the environment, and also facilitates a direct comparison with Ref.~\cite{WFL:09}, where the same model was used.

The single-axis error $E_{\mu }(\tau )$ will be dominated by the lowest
non-vanishing order of $\tau $, which we denote by $n_{\mu }$. I.e., $E_{\mu
}\sim \mathcal{O}(\tau ^{n_{\mu }})$ provided the first $n_{\mu }-1$ terms
of the power series expansion of $B_{\mu }^{{\prime \prime }}(\tau )$
vanish. More expliclity, we write 
\begin{equation}
B_{\mu }^{\prime \prime }(\tau )=\sum_{j=n_{\mu }}^{\infty }B_{\mu }^{\prime
\prime (j)}\tau ^{j},  \label{B_mu}
\end{equation}%
where 
\begin{equation}
B_{\mu }^{\prime \prime (j)}=\sum_{\vec{\alpha}_{j}^{\mu }}r_{\alpha
_{1}\alpha _{2}\ldots \alpha _{j}}^{\mu }\,\,B_{\alpha _{1}}B_{\alpha
_{2}}\cdots B_{\alpha _{j}},  \label{B_mu^j}
\end{equation}%
and where $\vec{\alpha}_{j}^{\mu }=\{\alpha _{1},...,\alpha _{j}\}$ such
that $\alpha _{j}\in \{1,x,y,z\}$ and $\mu =\prod_{j}\alpha _{j}$, with the
Pauli product rules $xy=z,zx=y,yz=x$. In this manner we have a convenient
notation to identify the environment operators present in each single-axis
error. For example, if $E_{z}(\tau )\sim \mathcal{O}(\tau ^{2})$ then the
possible summands comprising $B_{z}^{\prime \prime (2)}$ will be
proportional to $B_{0}B_{z}$, $B_{z}B_{0}$, $B_{x}B_{y}$, and $B_{y}B_{x}$.
The constituent operators, $B_{\alpha _{j}}$, are the environment operators
initially defined in Eq.~(\ref{eq:HSB}). The only terms of interest are
those proportional to $\tau ^{n_{\mu }}$, since these are the dominant terms
of $B_{\mu }^{\prime \prime }(\tau )$. Using Eqs.~(\ref{E_mu}), (\ref{B_mu})
and (\ref{B_mu^j}), and the submultiplicativity property of unitarily
invariant norms \cite{Bhatia:book}, the single-axis error is 
\begin{eqnarray}
E_{\mu }(\tau ) &\sim &\Vert B_{\mu }^{\prime \prime (n_{\mu })}\tau
^{n_{\mu }}\Vert _{F}  \notag \\
&\leq &\tau ^{n_{\mu }}\sum_{\vec{\alpha}_{n_{\mu }}^{\mu }}|r_{\alpha
_{1}\ldots \alpha _{n_{\mu }}}^{\mu }|\Vert B_{\alpha _{1}}\Vert _{F}\cdots
\Vert B_{\alpha _{n_{\mu }}}\Vert _{F}  \notag \\
&=&\tau ^{n_{\mu }}\sum_{\vec{\alpha}_{n_{\mu }}^{\mu }}\tilde{r}_{\alpha
_{1}\ldots \alpha _{n_{\mu }}}^{\mu }J_{\alpha _{1}}\cdots J_{\alpha
_{n_{\mu }}},
\end{eqnarray}%
where we have only kept the leading order contribution in $\tau $. The
coupling strength parameters defined in Eq.~(\ref{eq:J}) have been
incorporated into the sum, such that $J_{1}=\beta $ and $J_{x,y,z}\leq J$. The
parameters $\tilde{r}_{\alpha _{1}\alpha _{2}\ldots \alpha _{n_{\mu }}}^{\mu
}$ account for a conversion factor between the Frobenius and sup-operator
norms. Note that the reason we chose to work with the Frobenius norm is that  the distance measure (\ref{eq:Dist}) is expressed in terms of this norm. Factoring $J^{n_{\mu }}$ out
from the sum, the desired functional form of the single-axis error is 
\begin{equation}
\log (E_{\mu })\sim n_{\mu }\log (J\tau )+\log \left( \chi ^{\mu }\right) ,
\label{eq:logE}
\end{equation}%
with 
\begin{equation}
\chi ^{\mu }=\sum_{\vec{\alpha}_{n_{\mu }}^{\mu }}\tilde{r}_{\alpha
_{1}\alpha _{2}\ldots \alpha _{n_{\mu }}}^{\mu }\,\,\gamma _{\alpha
_{1}}\gamma _{\alpha _{2}}\cdots \gamma _{\alpha _{n_{\mu }}},
\end{equation}%
and with 
\begin{equation}
\gamma _{\alpha }=J_{\alpha }/J\leq 1.
\end{equation}

We have written the single-axis error as Eq.~(\ref{eq:logE}) in
anticipation of our numerical results, where we plot $E_{\mu }$ as a
function of the dimensionless parameter $J\tau $. In the $\beta \ll J$
regime it is $J$ which sets the relevant bath timesale and hence we expect
that $J\tau \lesssim 1$ should be a necessary condition for DD to be
beneficial over uncontrolled free evolution, and we shall see that our
simulations support this expectation. The quantity $\chi _{\mu }$ does not
depend on $\tau $ and hence will play the role of a constant offset. In the next section we shall unravel the connection between the suppression order $n_\mu$ and the sequence orders $N_1$ and $N_2$.

\section{Numerical Results}
\label{Numerics}

We now present a numerical analysis of QDD based on the single-axis errors
and the overall distance measure $D(U,G)$. The initial focus is the
single-axis error, which we use to quantify the decoherence suppression as a
function of inner and outer sequence orders.

\begin{figure}[t]
\includegraphics[width=\columnwidth]{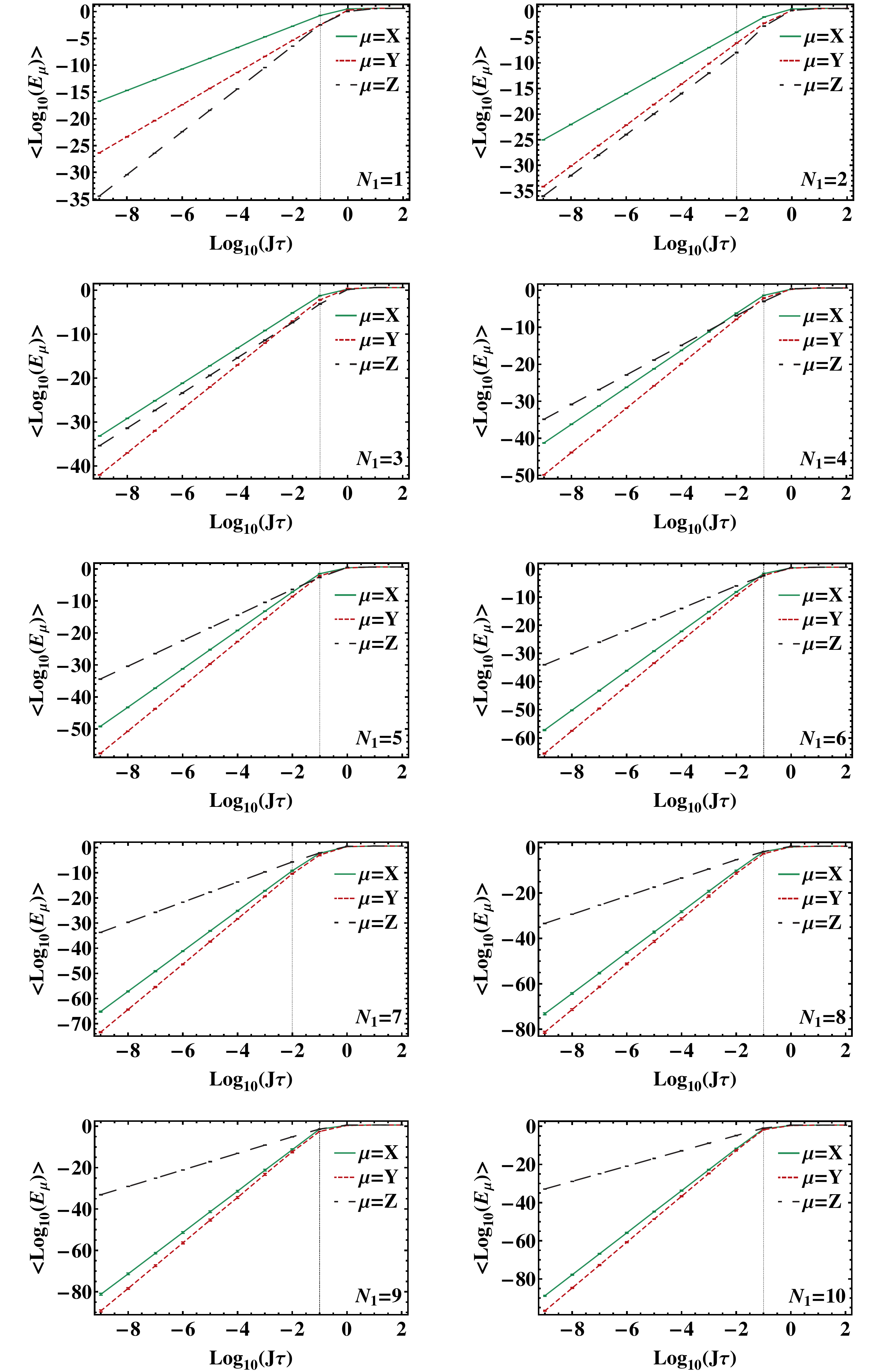}
\caption{(color online) Single-axis errors after one cycle of $U_{\Omega }^{(N_{1},N_2)}$ for outer sequence order $N_2=3$ and 
inner sequence orders $N_{1}=1,2,\ldots ,10$
as a function of $J\protect\tau $, averaged over $50$ random realizations of
the bath operators $B_{\protect\mu }$. Error bars are shown but are very small. In all our simulations we set $J=10^{-4}$ and $\beta=10^{-6}$. Single-axis error values were computed for $\log_{10}
(J\protect\tau )=-9,-8,\ldots ,2$. Lines are
guides to the eye. $E_{x}
$ is the solid green line,  $E_{y}$ is the dotted red line, and $E_{z}$ is
the dashed black line. Note that  $\log(E_{z})$ is the same in all six plots, with a slope of $N_2+1$. The slope of $\log(E_{x})$, on the other hand, is $N_1+1$. The slope of  $\log(E_{y})$ is $N_1+2$. Vertical lines denote the largest value of $J\tau$ utilized in the linear regression used to extract the slope $n_\mu$. 
}
\label{fig:N2=3Error}
\end{figure}

\begin{figure}[t]
\includegraphics[width=\columnwidth]{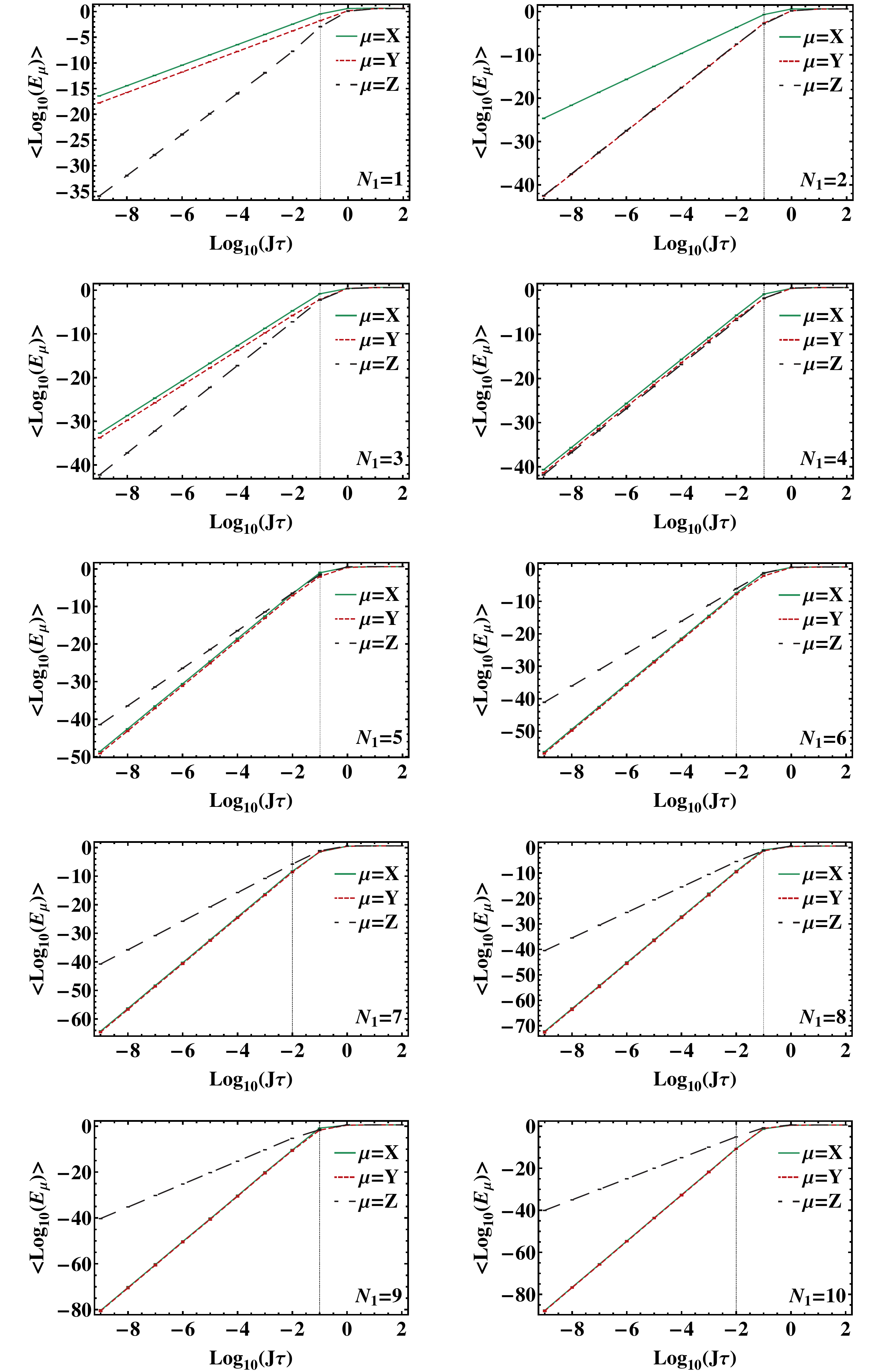}
\caption{(color online) Single-axis errors after one cycle of $U_{\Omega }^{(N_{1},N_2)}$ for outer sequence order $N_2=4$ and 
inner sequence orders $N_{1}=1,2,\ldots ,10$ (left to right, top to bottom)
as a function of $J\protect\tau $, averaged over $50$ random realizations of
the bath operators $B_{\protect\mu }$.  Other details as in Fig.~\protect\ref{fig:N2=3Error}, except that the single-axis error suppressed by both the inner and outer
sequence, $E_{y}(\protect\tau )$, exhibits a strong dependence on the parity
of the inner sequence.}
\label{fig:N2=4Error}
\end{figure}
\subsection{Single-axis errors}
Figures~\ref{fig:N2=3Error} and \ref{fig:N2=4Error} display the
single-axis errors as a function of $J\tau $ for $N_{2}=3$ and $N_{2}=4$,
respectively, with the inner sequence order varying from $N_{1}=1$ to $%
N_{1}=10$. The pure bath and system-bath interaction strengths were adjusted
such that the Hamiltonian is dominated by the interaction, $\beta \ll J$.
Each data point corresponds to a single cycle of the QDD sequence averaged
over $50$ random instances of the parameters $c_{\alpha \beta }^{\mu }$
appearing in Eq. (\ref{eq:B_mu}). Since we keep the minimum pulse interval
fixed, the total sequence duration increases with increasing $N_{1}$ and $N_{2}$. The
first thing to notice about Figures~(\ref{fig:N2=3Error}) and (\ref%
{fig:N2=4Error}) is that they match the prediction of Eq. (\ref{eq:logE})
very well in the regime of small $J\tau $. Namely, in all cases we observe a
constant slope, until $J\tau \sim \mathcal{O}(1)$. This is also in agreement with
the result of Ref. \cite{WFL:09}. 

A summary of the scalings for $E_{\mu }(J\tau )$ are given in Table~\ref{table:ErrorSlopes} for all combinations of $N_{1},N_{2} \in \{1,2,\ldots ,10\}$. The values of $n_\mu$ were extracted by performing linear regressions for $\log_{10} [E_{\mu }(J\tau )]$ between $\log_{10} (J\tau)=-9$ and the values of $\log_{10} (J\tau)$ indicated by the vertical lines in Figs.~\ref{fig:N2=3Error} and \ref{fig:N2=4Error}, and rounding to the nearest integer (in all cases the deviation from an integer value was at most in the third significant digit). We shall return to Table~\ref{table:ErrorSlopes} after presenting and discussing the data in the figures.

Let us then consider in detail the effect
of varying the inner and outer sequence orders on the single-axis errors. When $N_{2}>N_{1}$, higher order suppression is expected for the errors that
correspond to the system basis operators which anti-commute with the member
of the MOOS comprising the outer $X$-type sequence. Thus the single-axis
errors $E_{y}(\tau )$ and $E_{z}(\tau )$ are most heavily suppressed.

Since only the outer sequence can suppress $z$-axis, or $Z$-type errors [recall Eq.~(\ref{eq:QDD})], $E_{z}(J\tau )$ only gains additional error
suppression if the outer sequence order is increased. In Fig.~(\ref%
{fig:N2=3Error}), $N_{2}=3$ and $E_{z}(J\tau )\sim \mathcal{O}[(J\tau) ^{4}]$
for all $N_{1}$, exhibiting error suppression of the first $N_{2}$
terms of the interactions \ig{corresponding} \red{proportional} to $\sigma ^{z}$. Thus QDD operates with UDD
efficiency for error suppression by the outer nested sequence alone.

In a similar manner to $E_{z}(J\tau )$, the behavior of $E_{x}(J\tau )$ can be attributed to one of the two nested sequences. Namely, the error measured by $E_{x}$ is associated with $\sigma^{x}$, which only anti-commutes with the MOOS operator present in the inner $Z$-type
sequence. Determined
solely by the inner sequence order, $E_{x}(J\tau )\sim \mathcal{O}[(J\tau )^{N_{1}+1}]$. Essentially, the outer sequence has no effect on
the order of error suppression for $E_{x}(J\tau )$, as can be seen from Fig.~\ref{fig:N2=3Error} for $N_{2}=3$.

\begin{table*}[t]
(a)
$n_x = N_1+1$ for $N_1,N_2\in\{1,\dots,10\}$
\\ \vspace{0.5cm}
(b)
\begin{tabular}{c|c|c|c|c|c|c|c|c|c|c}
\multicolumn{11}{c}{$n_y$}\\ \hline
$N_1$ & $N_2=1$ & $N_2=2$ & $N_2=3$ & $N_2=4$ & $N_2=5$ & $N_2=6$ & $N_2=7$ & $N_2=8$ & $N_2=9$ & $N_2=10$\\
\hline
1& 3 & 2 & 3 & 2 & 3 & 2 & 3 & 2 & 3 & 2\\
2& 4 & 3 & 4 & 5 & 6 & 7 & 8 & 9 & 10& 11\\
3& 5 & 4 & 5 & 4 & 5 & 4 & 5 & 4 & 5 & 4\\
4& 6 & 5 & 6 & 5 & 6 & 7 & 8 & 9 & 10& 11\\
5& 7 & 6 & 7 & 6 & 7 & 6 & 7 & 6 & 7 & 6\\
6& 8 & 7 & 8 & 7 & 8 & 7 & 8 & 9 & 10& 11\\
7& 9 & 8 & 9 & 8 & 9 & 8 & 9 & 8 & 9 & 8\\
8& 10& 9 & 10& 9 & 10& 9 & 10& 9 & 10& 11\\
9& 11& 10& 11& 10& 11& 10& 11& 10& 11& 10\\
10&12& 11& 12& 11& 12& 11& 12& 11& 12& 11
\end{tabular}\\ \vspace{0.5cm}
(c)
\begin{tabular}{c|c|c|c|c|c|c|c|c|c|c}
\multicolumn{11}{c}{$n_z$}\\ \hline
$N_1$ & $N_2=1$ & $N_2=2$ & $N_2=3$ & $N_2=4$ & $N_2=5$ & $N_2=6$ & $N_2=7$ & $N_2=8$ & $N_2=9$ & $N_2=10$\\
\hline
1& 2 & 3 & 4 & 4 & 4 & 4 & 4 & 4 & 4 & 4\\
2& 2 & 3 & 4 & 5 & 6 & 7 & 8 & 9 & 10& 11\\
3& 2 & 3 & 4 & 5 & 6 & 7 & 8 & 8 & 8 & 8\\
4& 2 & 3 & 4 & 5 & 6 & 7 & 8 & 9 & 10& 11\\
5& 2 & 3 & 4 & 5 & 6 & 7 & 8 & 9 & 10& 11\\
6& 2 & 3 & 4 & 5 & 6 & 7 & 8 & 9 & 10& 11\\
7& 2 & 3 & 4 & 5 & 6 & 7 & 8 & 9 & 10& 11\\
8& 2 & 3 & 4 & 5 & 6 & 7 & 8 & 9 & 10& 11\\
9& 2 & 3 & 4 & 5 & 6 & 7 & 8 & 9 & 10& 11\\
10&2 & 3 & 4 & 5 & 6 & 7 & 8 & 9 & 10& 11
\end{tabular}
\caption{Summary of the scaling for all single-axis errors. Values of $n_\mu$ were extracted by performing a linear regression, rounded to the nearest integer, fitting the slopes of the straight line portions of the curves displayed in Figs.~\ref{fig:N2=3Error} and \ref{fig:N2=4Error}, and the additional Figs.~\ref{fig:N2=1Error}-\ref{fig:N2=6Error} in Appendix~\ref{app:A}, between $\log_{10} (J\tau)=-9$  and the values of $\log_{10} (J\tau)$ indicated by the vertical lines in these figures. (a) $E_x$, (b) $E_y$, and (c) $E_z$ for $N_{1},N_{2} \in \{1,2,\ldots ,10\}$. For $n_y$ and $n_z$ the outer sequence order $N_2$ is displayed in the top row and the inner sequence order $N_1$ in the first column. Each of the single-axis errors is dominated by the lowest order of $J\tau$, denoted $n_{\mu}$, therefore $E_{\mu}\sim\mathcal{O}[(J\tau)^{n_{\mu}}]$. Additional simulations (not shown) fully continue the trends seen in this table and summarized in Eqs.~\eqref{eq:nx}-\eqref{eq:ny} all the way up to $N_1,N_2\leq 24$.}
\label{table:ErrorSlopes}
\end{table*}
The interpretation for $E_{y}(J\tau )$ is not as simple, since this
single-axis error is compensated by both the inner and outer sequences. One might
expect both the inner and outer sequence to contribute to $Y$-type error suppression, i.e., $E_{y}(J\tau )$ to scale with $(J\tau) ^{\max (N_{1},N_{2})+1}$. 
However, if this were the case then, e.g., the case $N_{1}=1$ would display an equal order of
error suppression for both $E_{y}(\tau )$ and $E_{z}(\tau )$. Instead we find that $E_{y}(J\tau )\sim \mathcal{O}[(J\tau )^{N_{1}+2}]$ for $N_{2}=3$. Thus the 
suppression of $E_{y}(J\tau )$ is constrained by $N_{1}$, even when $N_{2}>N_{1}$, though it is larger by one order of magnitude than UDD error suppression efficiency for the inner sequence.

Similar observations apply for all \emph{odd}-order outer sequences
we have analyzed (see Appendix~\ref{app:A}, Figures~\ref{fig:N2=1Error} \ref{fig:N2=5Error}, \ref{fig:N2=7Error}, and \ref{fig:N2=9Error}).
Odd-order sequences are anti-symmetric with respect to time-reversal, and the conclusions concerning the case $N_2=3$ can
be generalized as follows: when the outer sequence is anti-symmetric, terms in the QDD evolution operator which anti-commute with only one element
of the MOOS are suppressed with UDD efficiency, determined by the order
of the nested sequence composed of the corresponding anti-commuting MOOS
operator (this applies to $\sigma^x$ and $\sigma^z$). In contrast, terma that anti-commutes with \emph{both}
elements of the MOOS are suppressed to one order beyond UDD efficiency,
dictated exclusively by the inner sequence order (this applies to $\sigma^y$). Below we will see how this observation is modified when we consider larger values of $N_2$.

Comparing the case of the anti-symmetric outer sequence of Fig.~\ref{fig:N2=3Error} to that of the symmetric sequence of $N_{2}=4$ in Fig.~\ref{fig:N2=4Error}, one notices immediately that there is a qualitative
difference. The single-axis error $E_{y}(J\tau )$, the component anti-commuting with both the inner and
outer sequences, $\sigma^y$, fluctuates strongly as a function
of $N_{1}$. A similar effect is observed for other even values of $N_2$ (see Appendix~\ref{app:A}, Figures~\ref{fig:N2=2Error}, \ref{fig:N2=6Error}, \ref{fig:N2=8Error}, and \ref{fig:N2=10Error}). Only the outer sequence order has been changed, therefore this
characteristic is entirely dependent on the fact that the outer sequence is
now symmetric.

Analogous to the anti-symmetric outer sequence, Fig.~\ref{fig:N2=4Error} shows that the single-axis error $E_{z}(J\tau )$ is independent of the
inner sequence order. The scaling $E_{z}(J\tau )\sim \mathcal{O}[(J\tau
)^{5}]$ holds for all $N_{1}$. Thus the single-axis error $E_{z}(J\tau )$ again
exhibits UDD efficiency, independent of the parity of the outer sequence. 
Similarly, again $E_{x}(J\tau )\sim \mathcal{O}[(J\tau)^{N_{1}+1}]$ in Fig.~\ref{fig:N2=4Error}. 

However, when we consider the $n_z$ results for all values of $N_1$ and $N_2$ we find that there are exceptions to this simple behavior. As can be seen from Table~\ref{table:ErrorSlopes}, when $N_1=1$ and $N_2\geq 4$, the value of $n_z$ is fixed at $4$. The same phenomenon is observed for $N_1=3$ and $N_2\geq 8$. \ig{An explanation of this anomaly is forthcoming \cite{KL:11}.}

On the basis of our numerical data we can summarize the scaling of the $X$ and $Z$-type single-axis errors as follows:
\begin{eqnarray}
n_x &=& N_1+1 ,
\label{eq:nx}
\end{eqnarray}
and
\begin{equation}
n_{z} = \left\{
 		    \begin{array}{ll}
 				N_2+1 &: N_1 \,\text{even}\\
 				N_2+1 &: N_1 \,\text{odd},\, N_2<2N_1+2\\
 				2N_1+2 &: N_1\,\text{odd},\, N_2\geq 2N_1+2
 		     \end{array}
 		\right.
\label{eq:nz}
\end{equation}

Qualitatively, we expect that when the inner sequence works imperfectly, as is the case for $N_1$ odd, the lowest order sequence will determine the scaling of the single axis error, and this is what is  stated in Eq.~(\ref{eq:nz}).

As is clear from Fig.~(\ref{fig:N2=4Error}), the scaling of $E_{y}(J\tau )$ is dependent on the parity of $N_{1}$. If the inner sequence
is of odd parity then $E_{y}(J\tau )\sim \mathcal{O}[(J\tau)^{N_{1}+2}]$ when $N_2$ is odd as well, or $E_{y}(J\tau )\sim \mathcal{O}[(J\tau)^{N_{1}+1}]$ when $N_2$ is even. Thus the scaling of $E_y$ is dominated by the inner sequence order when $N_1$ is odd. The situation changes when $N_1$ is even. Now, if $N_2$ is odd the sequence is still anti-symmetric, however there is an immediate improvement in error suppression, $E_{y}(J\tau )\sim \mathcal{O}[(J\tau)^{\max(N_{1}+1,N_{2})+1}]$. If
the complete sequence is fully symmetric (both $N_1$ and $N_2$ even) we also find a scaling dependent on both the inner and outer sequence orders, $E_{y}(J\tau )\sim \mathcal{O}[(J\tau)^{\max
(N_{1},N_{2})+1}]$.
We thus see that the suppression of interactions
which anti-commute with \emph{both} elements of the MOOS depends sensitively on the parity of the inner sequence order, and \ig{are} is summarized for $E_{y}(J\tau )\sim \mathcal{O}[(J\tau)^{n_{y}}]$ as

\begin{equation}
n_{y}=\left\{
		\begin{array}{ll}
		\max(N_1,N_2)+1 &: N_1 \,\text{even},\, N_2 \,\text{even}\\
		\max(N_1+1,N_2)+1 &: N_1 \,\text{even},\, N_2 \,\text{odd}\\
		N_1+1 &: N_1 \,\text{odd},\, N_2 \,\text{even}\\
		N_1+2 &: N_1 \,\text{odd},\, N_2 \,\text{odd}\\
		\end{array}
		\right.
\label{eq:ny}
\end{equation}

The dependence upon the symmetry of the inner
sequence, the parity of $N_{1}$, was first noted by Wang \& Liu in the context of 
overall QDD performance \cite{NUDD}.  The
dependence on the outer sequence symmetry, however, was not noted
previously. Our results show that the symmetry of the outer sequence impacts the efficiency of $\sigma^y$ error suppression as well.

The efficiency of QDD error suppression is directly related to the efficiency
of UDD. Each interaction that anti-commutes with at least one member of the
MOOS is expected to achieve UDD efficiency. Fully symmetric QDD, i.e., even 
order $N_{1}$ and $N_{2}$, recovers the efficiency of UDD for all single-axis 
errors. Consequently, QDD performs with optimal efficiency when it is
fully symmetric.

The interactions addressed only by the inner or outer
sequence are separately suppressed with UDD efficiency corresponding to the order of the corresponding
sequence performing the decoherence suppression. Equivalently, interactions
which anti-commute with only one member of the MOOS \ig{achieve} are suppressed with
UDD efficiency in the QDD scheme. On the other hand, error suppression of the interactions anti-commuting with both
elements of the MOOS is dependent upon the parity of both the inner and
outer sequence orders.

\subsection{Intermediate single-axis errors}

Rather than consider the single-axis errors just at the end of the QDD sequence, here \ig{There is another interesting way in which we can gain some additional insight into the dynamics of QDD. Namely, to understand the {\em intermediate} time-dependence of the errors,} we consider the single-axis errors prior to the application of each {$X$-type} outer sequence pulse. We will refer to these as ``intermediate single-axis errors'' since they are extracted during the QDD evolution, unlike those presented in Figures \ref{fig:N2=3Error}, \ref{fig:N2=4Error}, and \ref{fig:N2=1Error}-\ref{fig:N2=10Error} which are extracted at the end of the complete evolution. y studying this intermediate time-dependence of the errors we shall gain another interesting perspective on the manner in which the QDD sequence suppresses decoherence.

Let us define a set of ``intermediate QDD" sequences  as
\beq
\tilde{U}_{\Omega }^{(N_{1},j)} \equiv Z^{(N_{1})}(s_{j}\tau )XZ^{(N_{1})}(s_{j-1}\tau )\cdots XZ^{(N_{1})}(s_{1}\tau ),
\eeq
where $j \in \{1,\dots,N_2+1\}$. Thus, except for $j=1$, $\tilde{U}_{\Omega }^{(N_{1},j)}$ contains $j-1$ $X$-type pulses sandwiched between $j$ $Z$-type UDD sequences. When $j=1$ 
\beq
\tilde{U}_{\Omega }^{(N_{1},1)}\equiv Z^{(N_{1})}(s_{1}\tau )
\eeq 
is just the UDD sequence. We also separately define 
\beq
\tilde{U}_{\Omega }^{(N_{1},N_2+2)} \equiv {U}_{\Omega }^{(N_{1},N_2)} = X^{N_2} \tilde{U}_{\Omega }^{(N_{1},N_2+1)} ,
\eeq
i.e., the complete QDD sequence, Eq.~\eqref{eq:QDD}. Note that $\tilde{U}_{\Omega }^{(N_{1},N_2+2)}$ contains a final $X$ pulse if $N_2$ is odd, but not if $N_2$ is even.
Similarly to the error expansion~\eqref{eq:Uf}, we have the intermediate error expansion
\beq
\tilde{U}_{\Omega }^{(N_{1},j)}=B_{I}^{(j)}(\tau )+\sum_{\mu \in
\{x,y,z\}}\sigma ^{\mu }\otimes B_{\mu }^{(j) }(\tau ).
\label{tildeU1}
\eeq
In analogy to Eq.~\eqref{E_mu} we can now define the intermediate single-axis errors as
\begin{equation}
E_{\mu}^{(j)}(\tau)=\|B^{{(j)}}_{\mu}(\tau)\|_F,  
\label{E_mu^j}
\end{equation}
where $j\in \{1,N_2+2\}$. Note that for odd $N_2$ the errors $E_{\mu}^{(N_2+1)}$ and $E_{\mu}^{(N_2+2)}$ differ by a single instantaneous $X$ pulse (which is significant, as our simulations results will demonstrate), while for $N_2$ even $E_{\mu}^{(N_2+1)}= E_{\mu}^{(N_2+2)}$, so that below we do not plot $E_{\mu}^{(N_2+2)}$ in the even case.

\begin{figure}[t]
\includegraphics[width=\columnwidth]{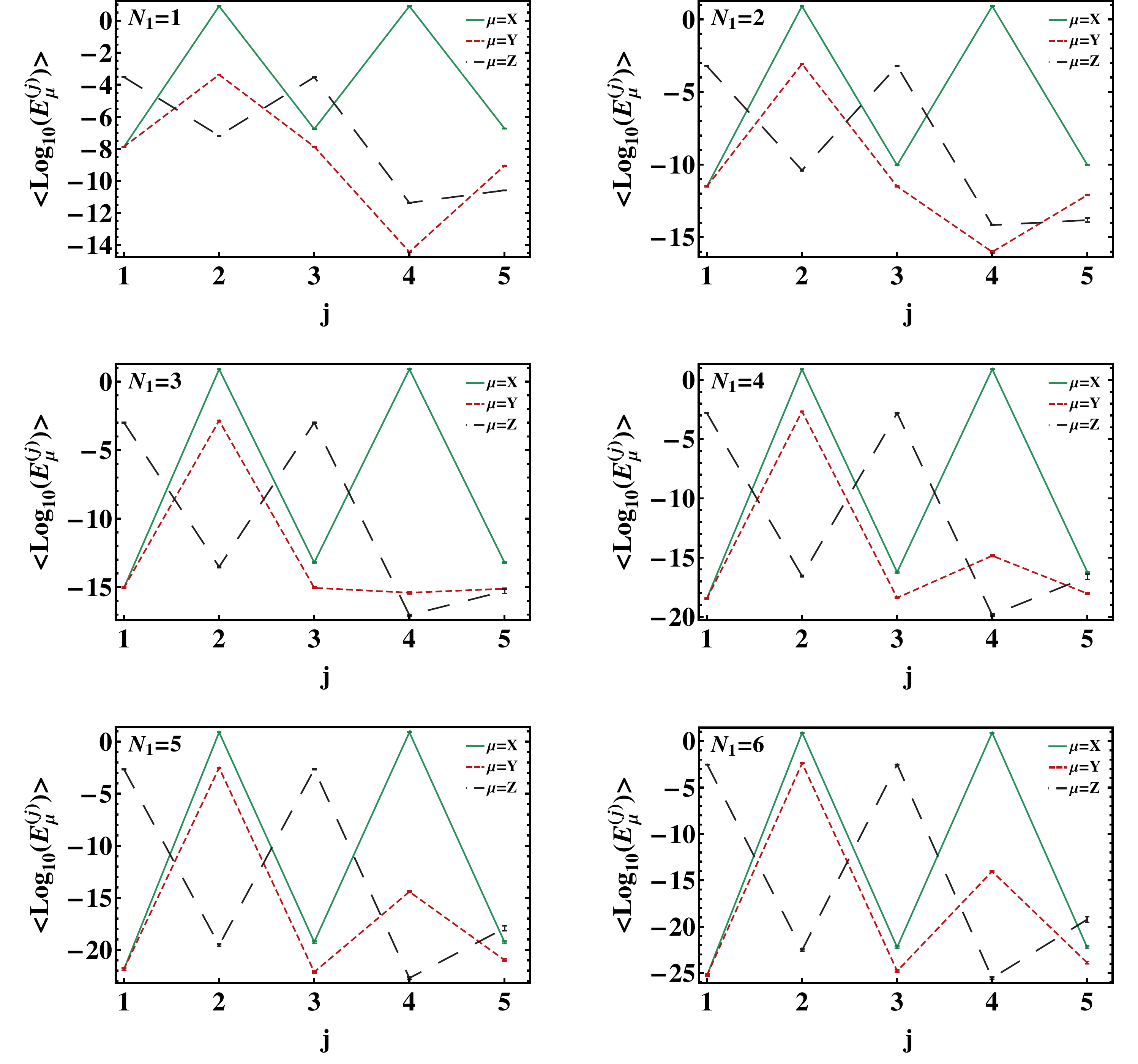}
\caption{Intermediate single-axis errors for $N_2=3$, as defined in Eq.~\eqref{E_mu^j}. For given $j$, the intermediate single-axis error is computed after $j$ innner $Z$-type UDD sequences separated by $j-1$ pulses. There are $N_2+1$ inner UDD sequences.  The point $j=5$ is the last $X$ pulse at the end of last inner UDD sequence, as required for odd $N_2$. Note that because the data points labeled $j=4$ and $j=5$ are separated by a single $X$ pulse, and our pulses are instantaneous, these points have no actual time delay between them.}
\label{fig:IntError3}
\end{figure}

Figures \ref{fig:IntError3} and \ref{fig:IntError4} display the intermediate single-axis errors for $N_2=3$ and $N_2=4$, respectively, with $N_1=1,2,\ldots,6$. The coupling parameters are fixed at $J=10^{-4}$ and $\beta=10^{-6}$ as in the previous figures. Additional results for odd $N_2$ are given in Appendix~\ref{app:A} in Figures \ref{fig:IntError1} and \ref{fig:IntError5}, and for even $N_2$ in Appendix~\ref{app:A} in Figures \ref{fig:IntError2} and \ref{fig:IntError6}.

{Several features are noteworthy in these figures. 

(i) $E_x^{(1)}$ and $E_y^{(1)}$ are equal and substantially smaller than $E_z^{(1)}$, and the difference grows as $N_1$ is increased. This is because the inner $Z$-type sequence only suppresses the $X$ and $Y$-type errors, and the point $j=1$ does not include the first $X$-type outer sequence pulse. Formally, this is expressed by
\bea
E_{\mu}^{(1)}(\tau)&=&\|B^{{(1)}}_{\mu}\|_F \sim (s_1 \tau)^{N_1+1},\quad \mu\in\{x,y\}\notag \\
E_{\nu}^{(1)}(\tau)&=& \|B^{{(1)}}_{\nu}\|_F \sim 1,\quad \nu\in\{I,z\} .
\label{E1}
\eea

(ii) The intermediate single-axis errors all fluctuate through{out} the QDD evolution. This is due to a reshuffling of the errors after each outer sequence $X$-type pulse is applied, a simple consequence of the rules of Pauli matrix multiplication. To see why in some detail, consider the effect of the first $X\equiv \sigma^x \otimes I$ pulse:
\bea
X\tilde{U}_{\Omega }^{(N_{1},1)}&=&\sigma ^{x}\otimes B_{I}^{(1)}+\sum_{\nu \in
\{x,y,z\}}\sigma ^{x}\sigma ^{\nu }\otimes B_{\nu }^{(1) }  \\
&=& B_{x}^{(1) } + \sigma ^{x}\otimes B_{I}^{(1)}+\sigma ^{y}\otimes B_{z}^{(1)}+\sigma ^{z}\otimes B_{y}^{(1)}, \notag
\eea
where we \ig{have absorbed} dropped factors of $i$\ig{ into the bath operators}. The reshuffling effect is clear: for example, the error single-axis $z$-type error now comes from $B_{y}^{(1)}$. To explain the $j=2$ behavior we should consider the effect of multiplying $X\tilde{U}_{\Omega }^{(N_{1},1)}$ by the next inner UDD sequence $Z^{(N_{1})}(s_{2}\tau )$. The $i$th inner UDD sequence has the expansion
\beq
Z^{(N_{1})}(s_{i}\tau ) = B_{I,i} + \sigma ^{x}\otimes B_{x,i}+\sigma ^{y}\otimes B_{y,i}+\sigma ^{z}\otimes B_{z,i} ,
\eeq
where similarly to Eq.~\eqref{E1} we have
\bea
\|B_{\mu,i}\|_F &\sim& (s_i \tau)^{N_1+1},\quad \mu\in\{x,y\}\notag \\
\|B_{\nu,i}\|_F &\sim& 1,\quad \nu\in\{I,z\} .
\label{Ei}
\eea
Using this to carry out the multiplication to the next order we have 

\bea
\tilde{U}_{\Omega }^{(N_{1},2)} &=& Z^{(N_{1})}(s_{2}\tau )X\tilde{U}_{\Omega }^{(N_{1},1)} \\
&=& [B_{I,2} + \sigma ^{x}\otimes B_{x,2}+\sigma ^{y}\otimes B_{y,2}+\sigma ^{z}\otimes B_{z,2}] \notag \\
&\times& [B_{x}^{(1)} + \sigma ^{x}\otimes B_{I}^{(1)}+\sigma ^{y}\otimes B_{z}^{(1)}+\sigma ^{z}\otimes B_{y}^{(1)}] \notag
\eea
Consequently 
\bea
E_{x}^{(2)}(\tau) \!\! &=& \!\! \|B_{I,2}B_{I}^{(1)} + B_{x,2}B_{x}^{(1)} + B_{y,2}B_{y}^{(1)} + B_{z,2}B_{z}^{(1)}\|_F \notag \\
\!\! &\sim& \!\! 1 \notag \\
E_{y}^{(2)}(\tau) \!\! &=& \!\! \|B_{I,2}B_{z}^{(1)} + B_{x,2}B_{y}^{(1)} + B_{y,2}B_{x}^{(1)} + B_{z,2}B_{I}^{(1)}\|_F \notag \\
\!\! &\sim& \!\! 1 \notag \\
E_{z}^{(2)}(\tau) \!\! &=& \!\! \|B_{I,2}B_{y}^{(1)} + B_{x,2}B_{z}^{(1)} + B_{y,2}B_{I}^{(1)} + B_{z,2}B_{x}^{(1)}\|_F \notag \\
\!\! &\sim& \!\! 2(s_1\tau)^{N_1+1} + 2(s_2\tau)^{N_1+1} ,
\eea
where $E_{x}^{(2)}(\tau)$ is dominated by $B_{I,2}B_{I}^{(1)}$ and $E_{y}^{(2)}(\tau)$ is dominated by $B_{I,2}B_{z}^{(1)}$, neither of which is suppressed, whence the $\sim 1$ result. On the other hand every one of the terms in $E_{z}^{(2)}(\tau)$ is suppressed. Hence, as can be seen in Figures \ref{fig:IntError3} and \ref{fig:IntError4} (and their companions, Figures~\ref{fig:IntError1}-\ref{fig:IntError6} in the Appendix), at $j=2$ both the $X$ and $Y$-type errors have increased relative to $j=1$, while the $Z$-type error has decreased. One can similarly understand the remaining oscillations of the intermediate single-axis errors in terms of this reshuffling of error types.

(iii) $E_x$ and $E_z$ oscillate out of phase, while $E_y$ oscillates in phase with $E_x$ for even $N_2$, but not necessarily for odd $N_1$. This is again a consequence of error reshuffling. The $Y$-type error behaves differently from the other two since it experiences suppression from both the inner and outer sequences. For the same reason we always find $E_y^{(j)} < E_x^{(j)}$. 

(iv) $E_x$ attains its minimum for $j=1$ and then slowly increases, though while maintaining its suppression \emph{order}. This is because the $X$-type error is \ig{essentially} suppressed only by the inner sequences, and these are simply applied to it with fixed order ($N_1$), a total of $N_2$ or $N_2+1$ times. Repeated application of the inner UDD sequence is similar to the periodic DD (PDD) protocol, \ig{which} whose performance is well known to deteriorate as time grows \cite{KhodjastehLidar:07,KhodjastehLidar:08}. The reason is that the error accumulates over time, without a mechanism for reducing it.

(v) There does not appear to be much of a difference between even and odd values of $N_2$ in terms of the intermediate single-axis errors. One difference is that $E_y^{(j)}$ tends to be more erratic for odd $N_2$ at high $j$ values. We do not have a simple explanation for this behavior. Another difference is that for even $N_2$ all single-axis errors have the same final value when $N_1=N_2$, but for odd $N_2$ the $X$-type error is always slightly worse at the end of the sequence, thus setting the bottleneck. Perhaps additional pulse interval optimization can remove this asymmetry.

(vi) Only at the very end are all three single-axis errors simultaneously small. Thus, while suppression of one error type can be achieved in the middle of the QDD sequence, one must wait until its completion to suppress all errors.

\begin{figure}[t]
\includegraphics[width=\columnwidth]{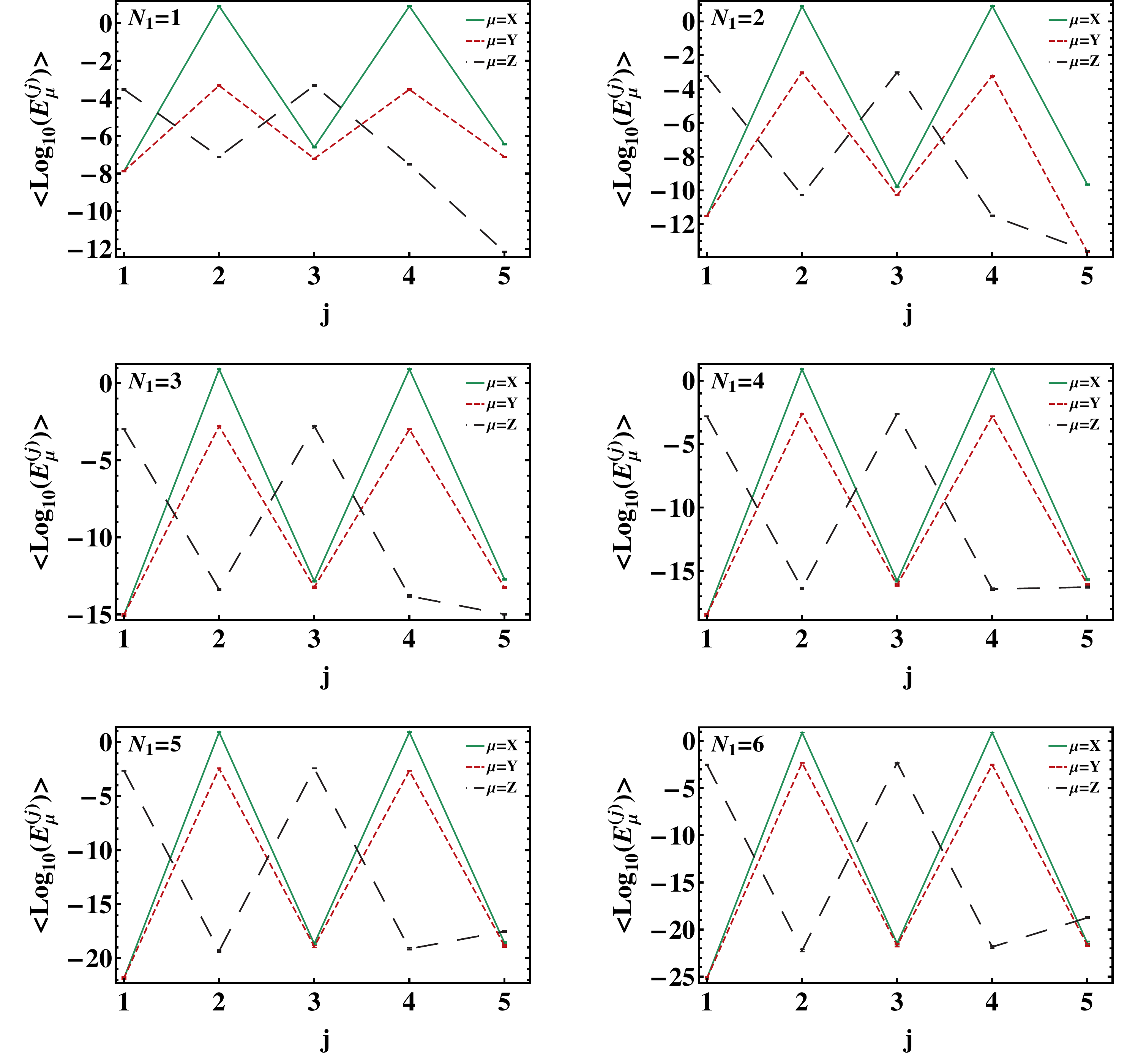}
\caption{(color online) 
Intermediate single-axis errors for $N_2=4$. As in Fig.~\ref{fig:IntError3} except that there is no final $X$ pulse for $N_2$ even, i.e., there are $N_2+1$ inner UDD sequences separated by $N_2$ $X$ pulses.}
\label{fig:IntError4}
\end{figure}

\subsection{Overall performance}

\begin{table*}[t]
\begin{tabular}{c|c|c|c|c|c|c|c|c|c|c}
\multicolumn{11}{c}{$n_D$}\\ \hline
$N_1$ & $N_2=1$ & $N_2=2$ & $N_2=3$ & $N_2=4$ & $N_2=5$ & $N_2=6$ & $N_2=7$ & $N_2=8$ & $N_2=9$ & $N_2=10$\\
\hline
1& 2 & 2 & 2 & 2 & 2 & 2 & 2 & 2 & 2 & 2\\
2& 2 & 3 & 3 & 3 & 3 & 3 & 3 & 3 & 3 & 3\\
3& 2 & 3 & 4 & 4 & 4 & 4 & 4 & 4 & 4 & 4\\
4& 2 & 3 & 4 & 5 & 5 & 5 & 5 & 5 & 5 & 5\\
5& 2 & 3 & 4 & 5 & 6 & 6 & 6 & 6 & 6 & 6\\
6& 2 & 3 & 4 & 5 & 6 & 7 & 7 & 7 & 7 & 7\\
7& 2 & 3 & 4 & 5 & 6 & 7 & 8 & 8 & 8 & 8\\
8& 2 & 3 & 4 & 5 & 6 & 7 & 8 & 9 & 9 & 9\\
9& 2 & 3 & 4 & 5 & 6 & 7 & 8 & 9 & 10& 10\\
10&2 & 3 & 4 & 5 & 6 & 7 & 8 & 9 & 10& 11
\end{tabular}
\caption{Summary of the scaling of the overall distance measure $D$ with respect to inner and outer sequence orders, $N_1$ and $N_2$, respectively. Values of $n_D$ were extracted by performing a linear regression, rounded to the nearest integer, fitting the slopes of the straight line portions of the curves displayed in Figs.~\ref{fig:N2=3Error} and \ref{fig:N2=4Error} and the additional Figs.~\ref{fig:N2=1dist}-\ref{fig:N2=6dist} in Appendix \ref{app:A} between $\log(J\tau)=-9$ and the values of $\log(J\tau)$ indicated by the vertical lines in these figures. 
The outer sequence order $N_2$ is displayed in the top row and the inner sequence order $N_1$ in the first column. We find that, as expected, $n_D = \min(n_x,n_y,n_z)$. Additional simulations (not shown) fully continue the trends seen in this table and summarized in Eq.~\eqref{eq:nD} all the way up to $N_1,N_2\leq 24$.}
\label{table:DSlopes}
\end{table*}

While the single-axis error analysis presented \ig{above }in the previuous two subsections helps in unravelling the mechanism of QDD performance, it does of course not tell the whole story. We now present our results for the distance measure $D(U,I)$ [Eq.~\eqref{eq:Dist}], which \ig{encompasses }provides a complete quantitative description of QDD performance. We expect this overall performance of QDD to be
dictated by the lowest order of $\tau $ present in the final evolution
operator, Eq.~(\ref{eq:Uf}), i.e.,
\begin{equation}
D \sim \mathcal{O}[(J\tau)^{n_D}]
\end{equation}
where 
\begin{equation}
n_D = \min (n_{x},n_{y},n_{z}).
\label{eq:nD}
\end{equation}

\begin{figure}[t]
\includegraphics[width=\columnwidth]{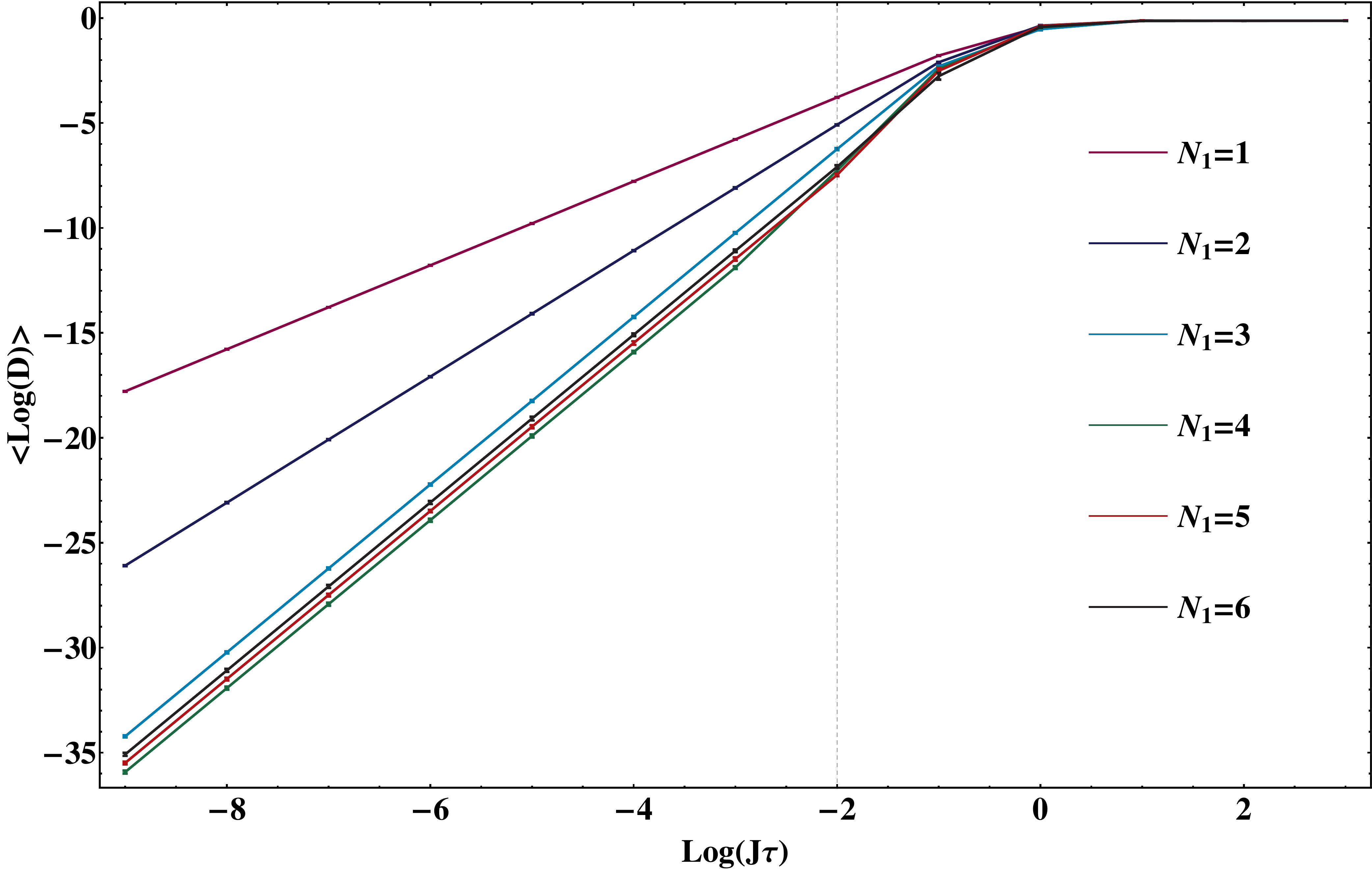}
\caption{(color online) Overall QDD distance measure after one cycle of $U_{\Omega }^{(N_{1},N_2)}$ for outer sequence order $N_2=3$ and 
inner sequence orders $N_{1}=1,2,\ldots ,6$,
as a function of $J\protect\tau $, averaged over $50$ random realizations of
the bath operators $B_{\protect\mu }$.  The performance of QDD progressively improves with increasing $%
N_1$ up to $N_2=N_1$, indicating that $\min\{N_1,N_2\}$ dominates QDD performance.}
\label{fig:N2=3F}
\end{figure}

\begin{figure}[t]
\includegraphics[width=\columnwidth]{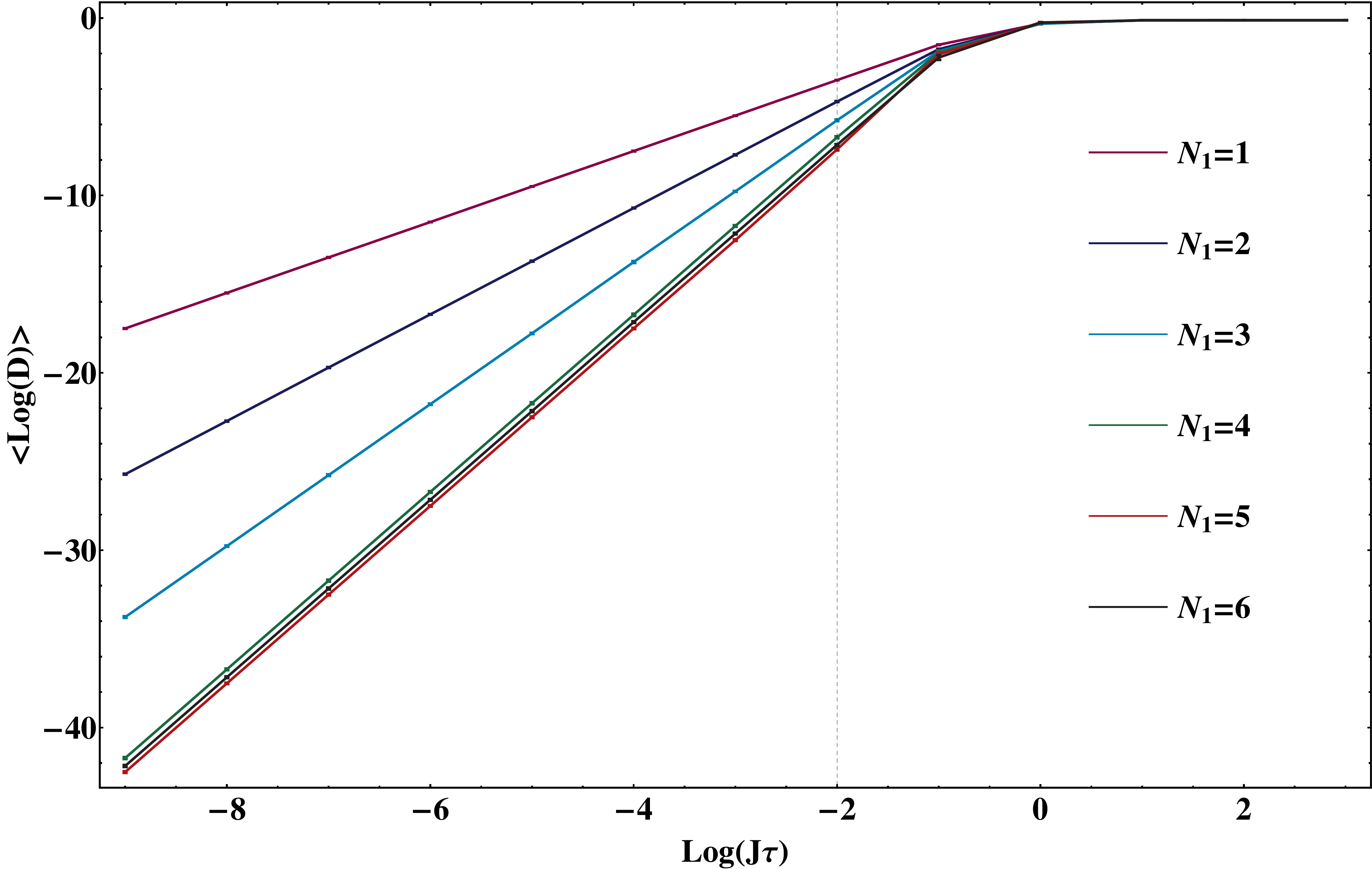}
\caption{(color online) Overall QDD distance measure after one cycle of $U_{\Omega }^{(N_{1},N_2)}$ for outer sequence order $N_2=4$ and 
inner sequence orders $N_{1}=1,2,\ldots ,6$,
as a function of $J\protect\tau $, averaged over $50$ random realizations of
the bath operators $B_{\protect\mu }$.  
The dependence of the order of error suppression on $\min\{N_1,N_2\}$ is again observed.}
\label{fig:N2=4F}
\end{figure}

Overall QDD performance for $%
N_2=3$ and $N_2=4$ is shown in Figures~\ref{fig:N2=3F} and \ref{fig:N2=4F}, respectively. The outer sequence order $N_2$ is fixed and the
inner sequence order $N_1$ is varied from $1$ to $6$. These results are for the same model considered in the previous subsection.
Additional results are given in Appendix~\ref{app:A} for $N_2=1,2,5,6$ (see Figures~\ref{fig:N2=1dist}-\ref{fig:N2=6dist}). A summary of the distance
scaling results is presented in Table~\ref{table:DSlopes}.

Considering $N_2=3$ first (Fig.~\ref{fig:N2=3F}), when $N_2>N_1$ the
overall order of error suppression is hindered by the inner sequence
order. This is evident by the increasing order of error suppression as 
$N_1$ increases. In this regime the lower sequence order is determined by the inner
sequence, therefore the scaling of $D$ is equivalent to that of $E_x$, i.e., $D\sim%
\mathcal{O}[(J\tau)^{N_1+1}]$. As $N_1$ passes $N_2$, there is a saturation of
error suppression corresponding to a performance bounded by the lower
outer sequence order. The amplitude of performance increases slightly beyond 
$N_2=N_1$, however begins to decrease when $N_1>N_2$, as evidenced not by the slope but by the offset of the distance curves. Namely, the ordering, from worst to best, is $N_1=6,5,4$. The latter is an
interesting feature not easily deduced from the single-axis errors. Increasing the
inner sequence order results in an accumulation of error for the single-axis
error dominating the performance; when $N_1>N_2+1$ this
corresponds to $E_z$.

The results are similar for $N_2=4$, as shown by Fig.~\ref{fig:N2=4F}. The order of error suppression, given by the slope, increases until 
$N_1=N_2$ in correspondence with an overall performance dominated by the
lowest order of $\tau$ present in $E_{\mu}$. In addition to the saturation
of the order of error suppression, we again observe an offset-related deterioration. Namely, $N_1=6$ is slightly worse than $N_1=5$.

\section{Conclusions}
\label{Conclusion}

This work presents a comprehensive numerical analysis of the error suppression characteristics of QDD. This was achieved by isolating the single-axis errors associated with each system basis operator in the system-bath interaction. The order of error
suppression was determined by computing the single-axis error as a function
of the minimum pulse interval. We performed our analysis for a model in which the system-environment interaction dominated the internal bath dynamics, so that we could study the properties of the single-axis
errors in the regime where DD is most beneficial. We constructed our QDD sequences with $N_1$ $Z$-type pulses comprising the inner sequence, and $N_2$ $X$-type pulses comprising the outer sequence. We found that the system-bath interaction term proportional to $\sigma^x$ \ig{was }is suppressed with UDD efficiency 
for all values of \ig{the inner and outer sequence orders} $N_{1}$ and $N_{2}$ [Eq.~(\ref{eq:nx})]. \ig{Both} The interactions proportional to $\sigma^z$ and $\sigma^y$ both \ig{interactions }exhibit parity effects [Eqs.~(\ref{eq:nz}), (\ref{eq:ny})] whose origins are the symmetry or anti-symmetry of the inner and outer UDD sequences. Of course, permuting the pulse types of the inner and outer sequences will correspondingly modify these conclusions.

We also performed an analysis of the intermediate time-dependent performance of QDD. We found that the single-axis errors are strongly time-dependent, oscillating between outer-sequence pulses, until they all converge to nearly the same value after the final outer-sequence pulse. The closest convergence occurs for QDD sequences with equal inner and outer orders. \ig{The strong time-dependence suggests that it can be advantageous to interrupt a QDD sequence before it has completed if the goal is to suppress a particular error type, but not all errors.}

Finally, we computed the overall performance of
QDD using an appropriate distance measure, and reconciled its scaling with that of the single-axis errors. We showed that overall QDD error suppression scales with the lowest order of single-axis error suppression, i.e., the first non-vanishing contribution appears at order $\min(N_1,N_2)+1$. QDD accomplishes this by applying $(N_1+1)(N_2+1)$ pulses. We conjecture that similarly, for NUDD with $K$ nested UDD sequences using $(N_1+1)(N_2+1)\cdots (N_K+1)$ pulses, the first non-vanishing contribution will appear at order $\min_j(N_j)+1$.

In this work we treated the pulses as ideal, instantaneous operations. However, this is of course an idealization. An important topic for future study is robustness with respect to pulse errors, whether random or systematic. This topic has been addressed for UDD both theoretically \cite{Uhrig:09a,Pasini:10} and experimentally \cite{PhysRevA.83.032303}, and the overall conclusion is that pulse errors can have a dramatic negative impact unless they are compensated for. Some combination of pulse shaping and optimization will surely be required to overcome this problem in the context of QDD as well.

\begin{acknowledgments}
We are grateful to Gonzalo Alvarez, Wan-Jung Kuo,
Stefano Pasini, Dieter Suter, and G\"{o}tz Uhrig for very helpful
discussions. DAL acknowledges support from the U.S. Department of Defense
and the NSF under Grants No. CHM-1037992 and CHM-924318.
\end{acknowledgments}

\appendix

\section{Additional Numerical Results}
\label{app:A}

In this appendix we present additional figures in support of the numerical results presented in the body of the paper.

\begin{figure}[H]
\includegraphics[width=\columnwidth]{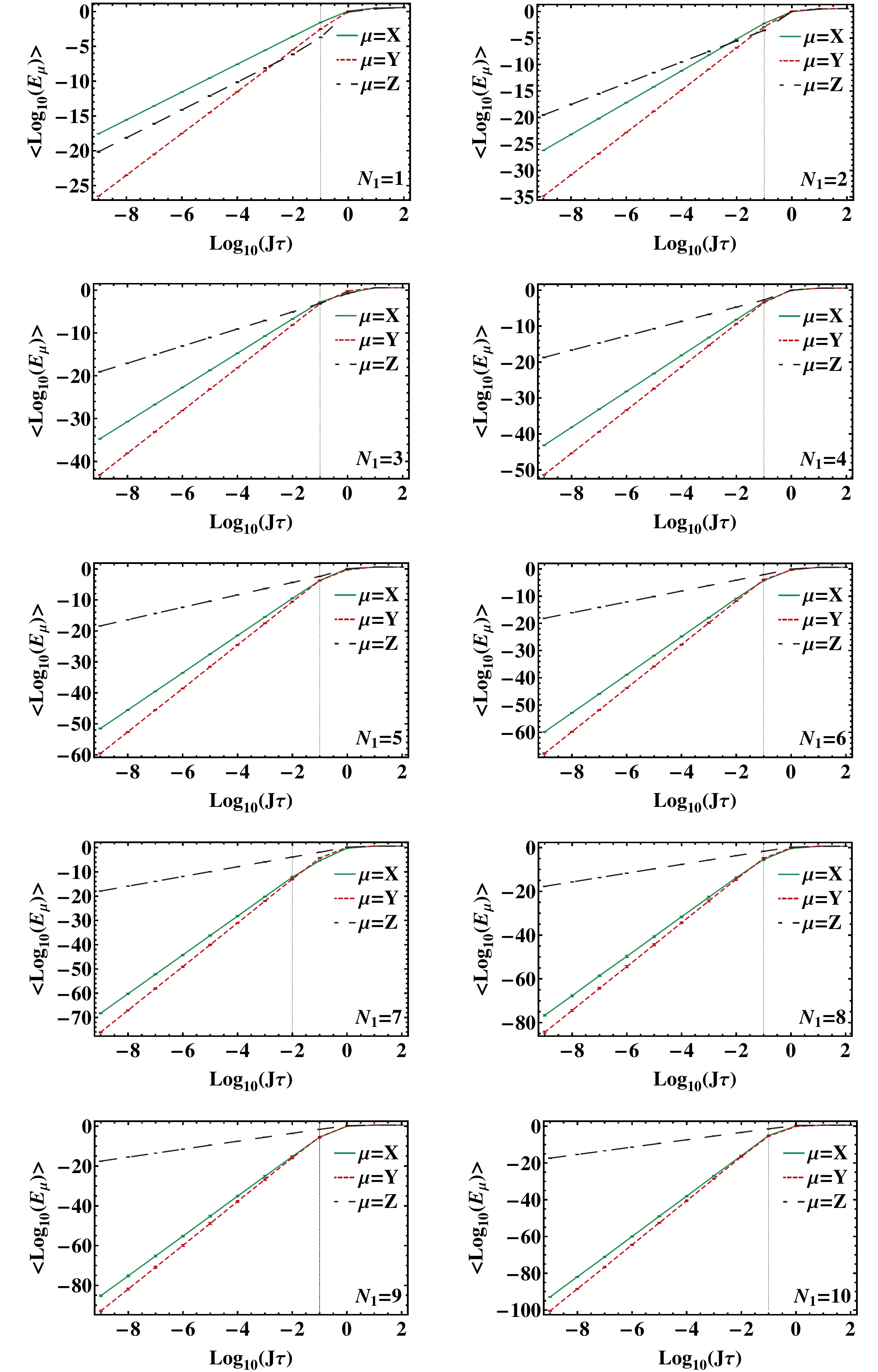} 
\caption{(color online) Single-axis errors after one cycle for $N_2=1$ and $%
N_1=1,2,\ldots,10 $ (left to right, top to bottom). See Fig.~\ref{fig:N2=3Error} for additional details.}
\label{fig:N2=1Error}
\end{figure}

\begin{figure}[H]
\includegraphics[width=\columnwidth]{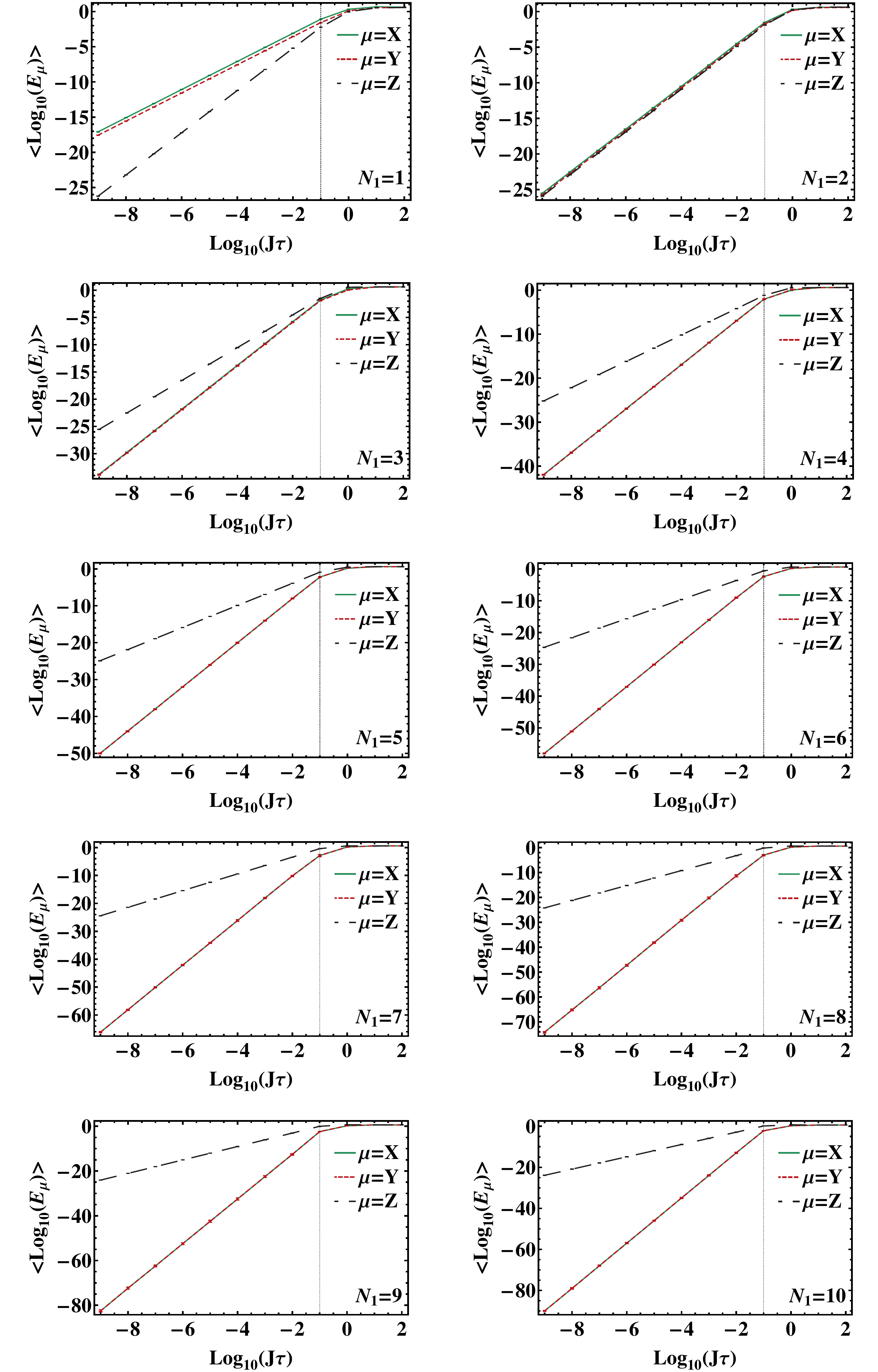}
\caption{(color online) Single-axis errors after one cycle for $N_2=2$ and $%
N_1=1,2,\ldots,10 $ (left to right, top to bottom). See Fig.~\ref{fig:N2=4Error} for additional details.}
\label{fig:N2=2Error}
\end{figure}

\begin{figure}[H]
\includegraphics[width=\columnwidth]{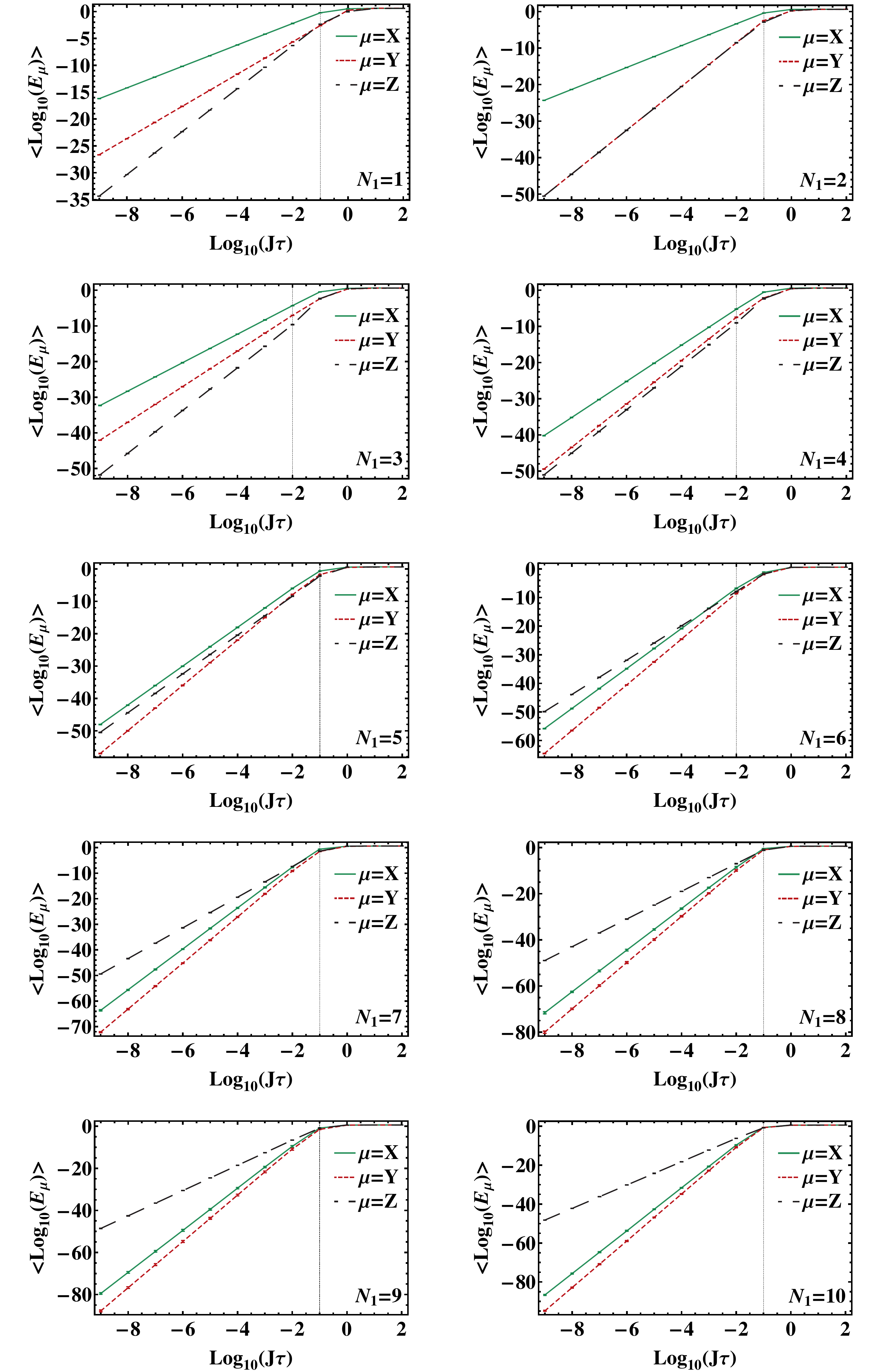}
\caption{(color online) Single-axis errors after one cycle for $N_2=5$ and $%
N_1=1,2,\ldots,10 $ (left to right, top to bottom). See Fig.~\ref{fig:N2=3Error} for additional details.}
\label{fig:N2=5Error}
\end{figure}

\begin{figure}[H]
\includegraphics[width=\columnwidth]{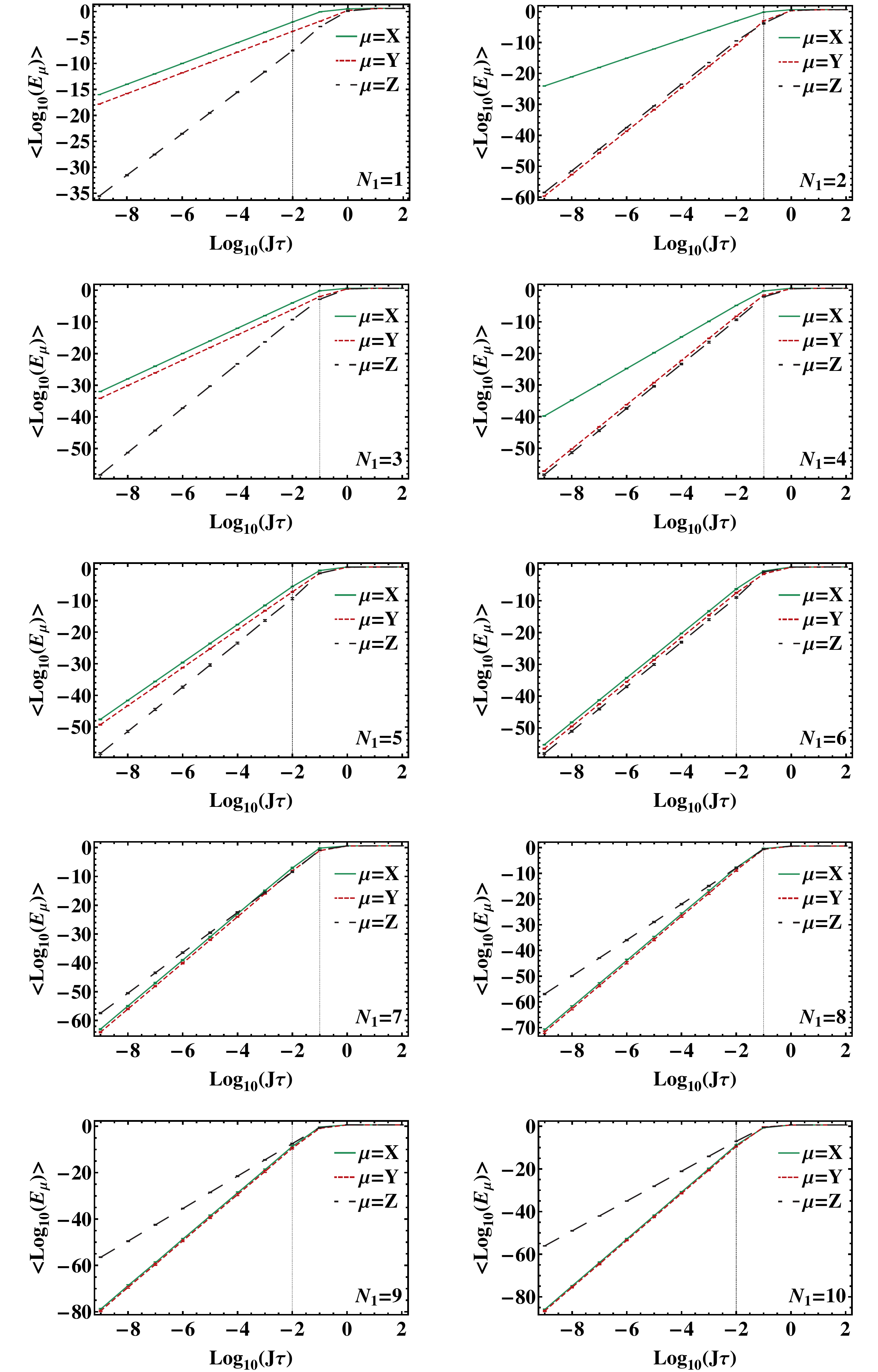}
\caption{(color online) Single-axis errors after one cycle for $N_2=6$ and $%
N_1=1,2,\ldots,10 $ (left to right, top to bottom). See Fig.~\ref{fig:N2=4Error} for additional details.}
\label{fig:N2=6Error}
\end{figure}

\begin{figure}[H]
\includegraphics[width=\columnwidth]{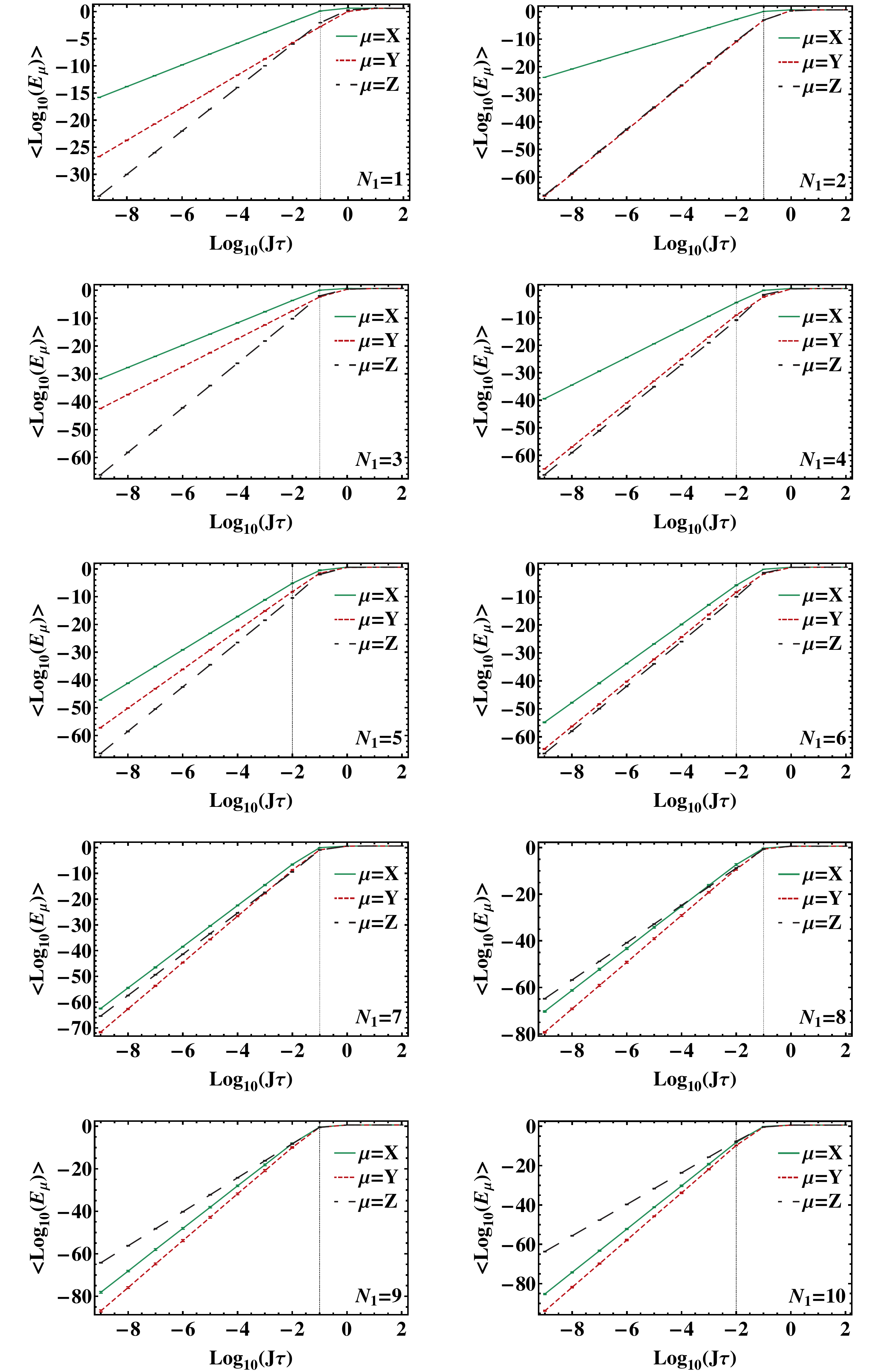}
\caption{(color online) Single-axis errors after one cycle for $N_2=7$ and $%
N_1=1,2,\ldots,10 $ (left to right, top to bottom). See Fig.~\ref{fig:N2=4Error} for additional details.}
\label{fig:N2=7Error}
\end{figure}

\begin{figure}[H]
\includegraphics[width=\columnwidth]{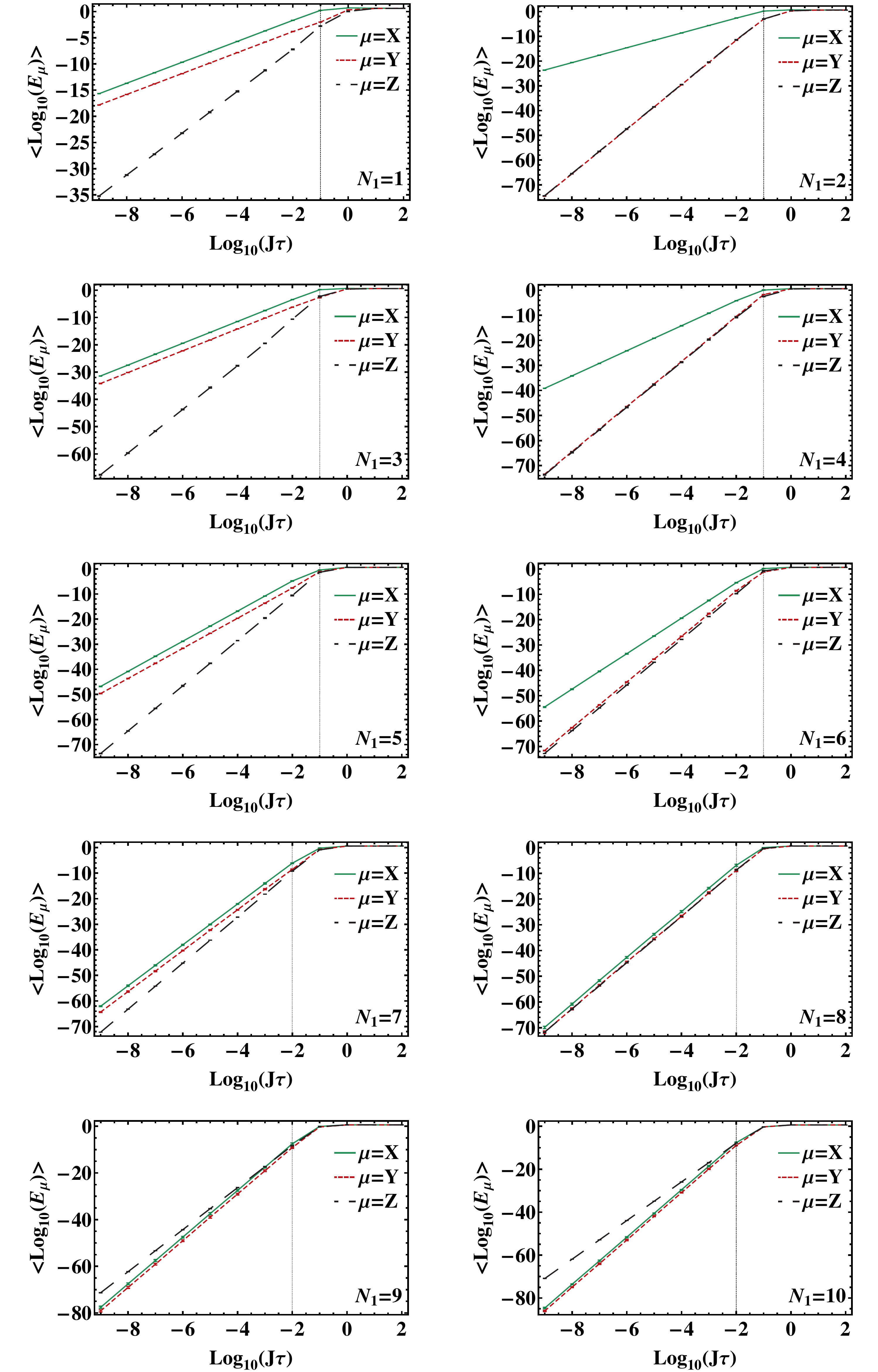}
\caption{(color online) Single-axis errors after one cycle for $N_2=8$ and $%
N_1=1,2,\ldots,10 $ (left to right, top to bottom). See Fig.~\ref{fig:N2=4Error} for additional details.}
\label{fig:N2=8Error}
\end{figure}

\begin{figure}[H]
\includegraphics[width=\columnwidth]{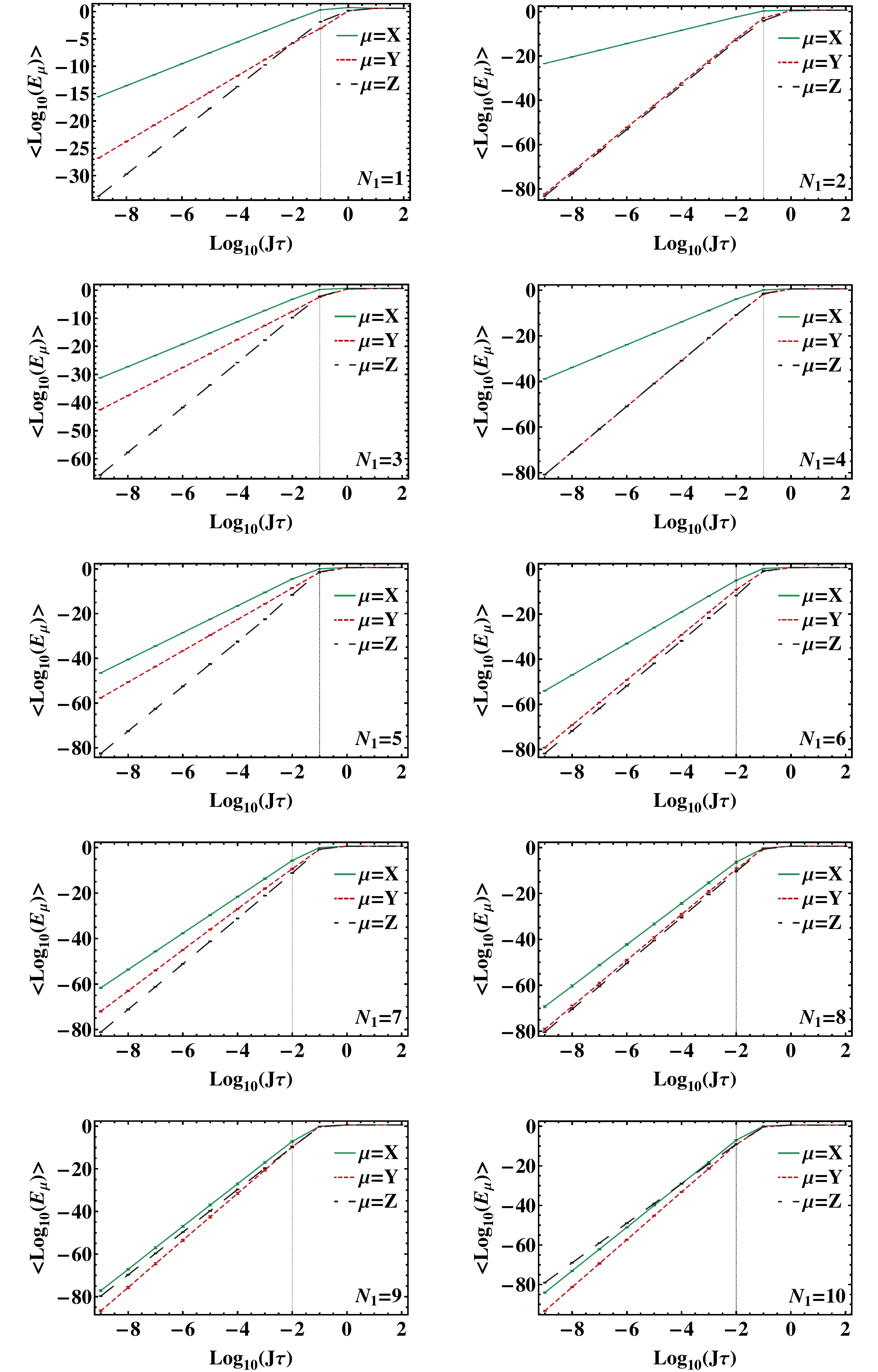}
\caption{(color online) Single-axis errors after one cycle for $N_2=9$ and $%
N_1=1,2,\ldots,10 $ (left to right, top to bottom). See Fig.~\ref{fig:N2=4Error} for additional details.}
\label{fig:N2=9Error}
\end{figure}

\begin{figure}[H]
\includegraphics[width=\columnwidth]{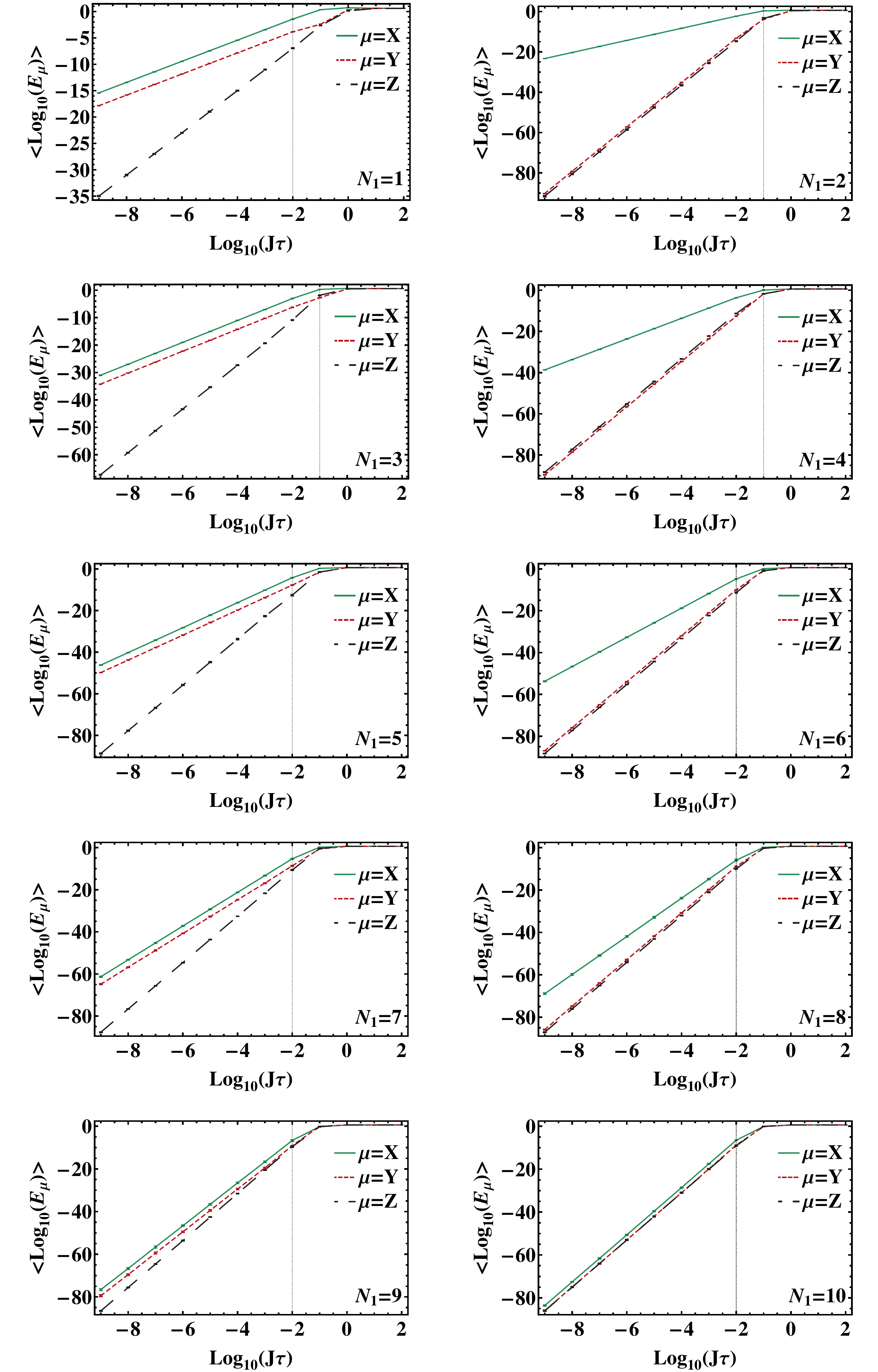}
\caption{(color online) Single-axis errors after one cycle for $N_2=10$ and $%
N_1=1,2,\ldots,10 $ (left to right, top to bottom). See Fig.~\ref{fig:N2=4Error} for additional details.}
\label{fig:N2=10Error}
\end{figure}

\begin{figure}[H]
\includegraphics[width=\columnwidth]{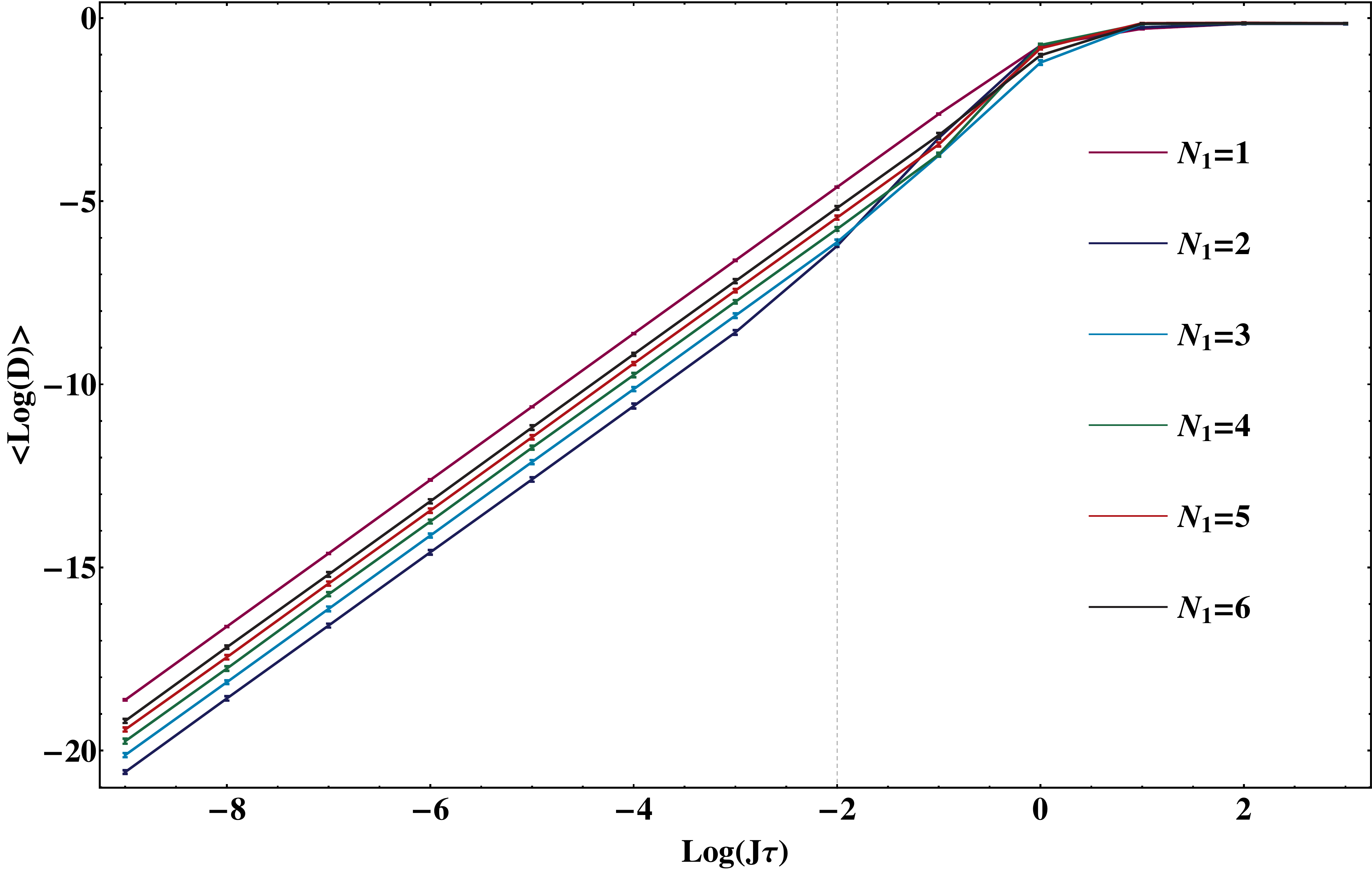}
\caption{(color online) Overall QDD distance measure after one cycle for $N_2=1$ and $%
N_1=1,2,\ldots,6 $ (left to right, top to bottom). See Fig.~\ref{fig:N2=3F} for additional details.}
\label{fig:N2=1dist}
\end{figure}

\begin{figure}[H]
\includegraphics[width=\columnwidth]{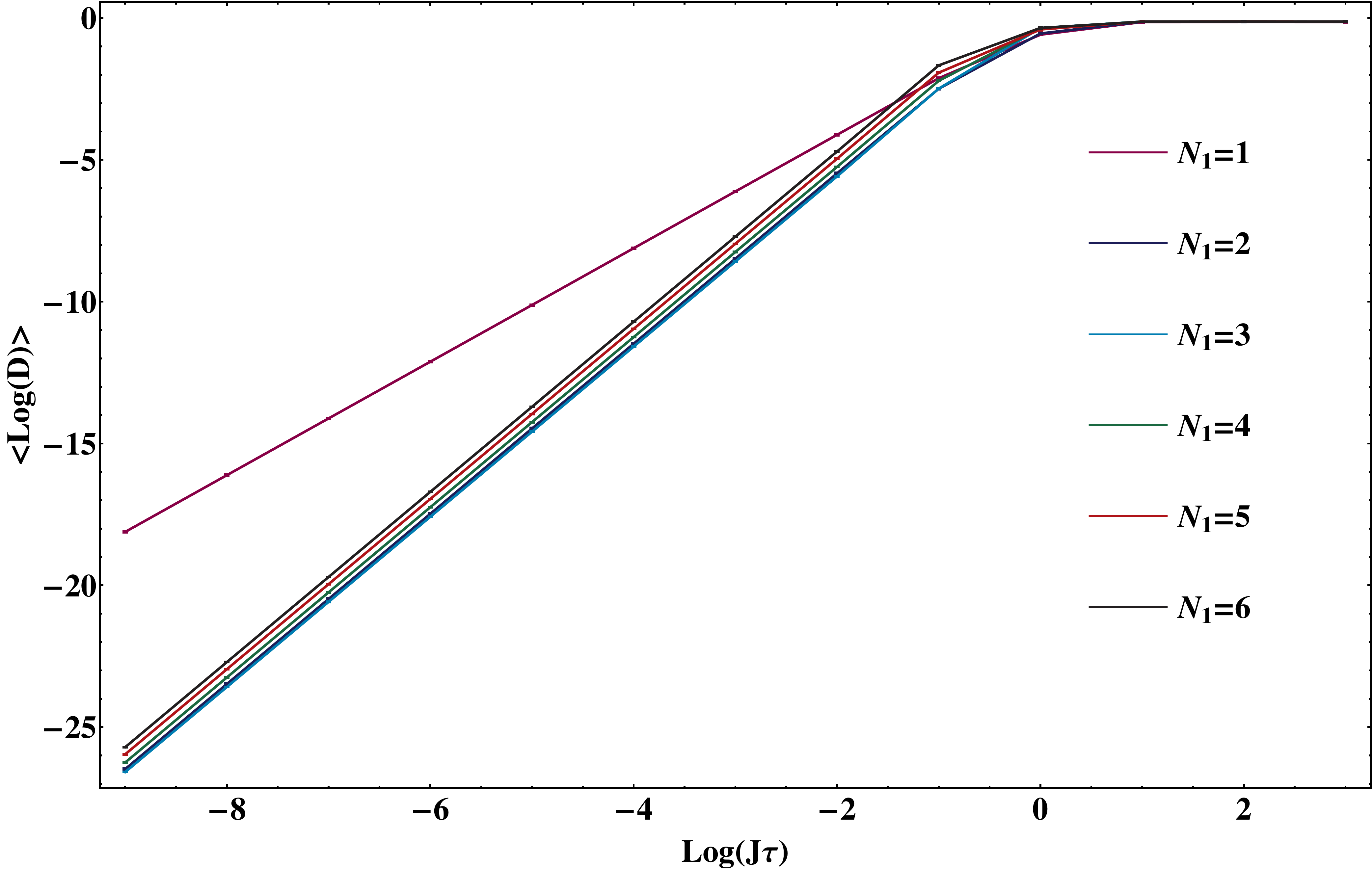}
\caption{(color online) Overall QDD distance measure after one cycle for $N_2=2$ and $%
N_1=1,2,\ldots,6 $ (left to right, top to bottom). See Fig.~\ref{fig:N2=4F} for additional details.}
\label{fig:N2=2dist}
\end{figure}

\begin{figure}[H]
\includegraphics[width=\columnwidth]{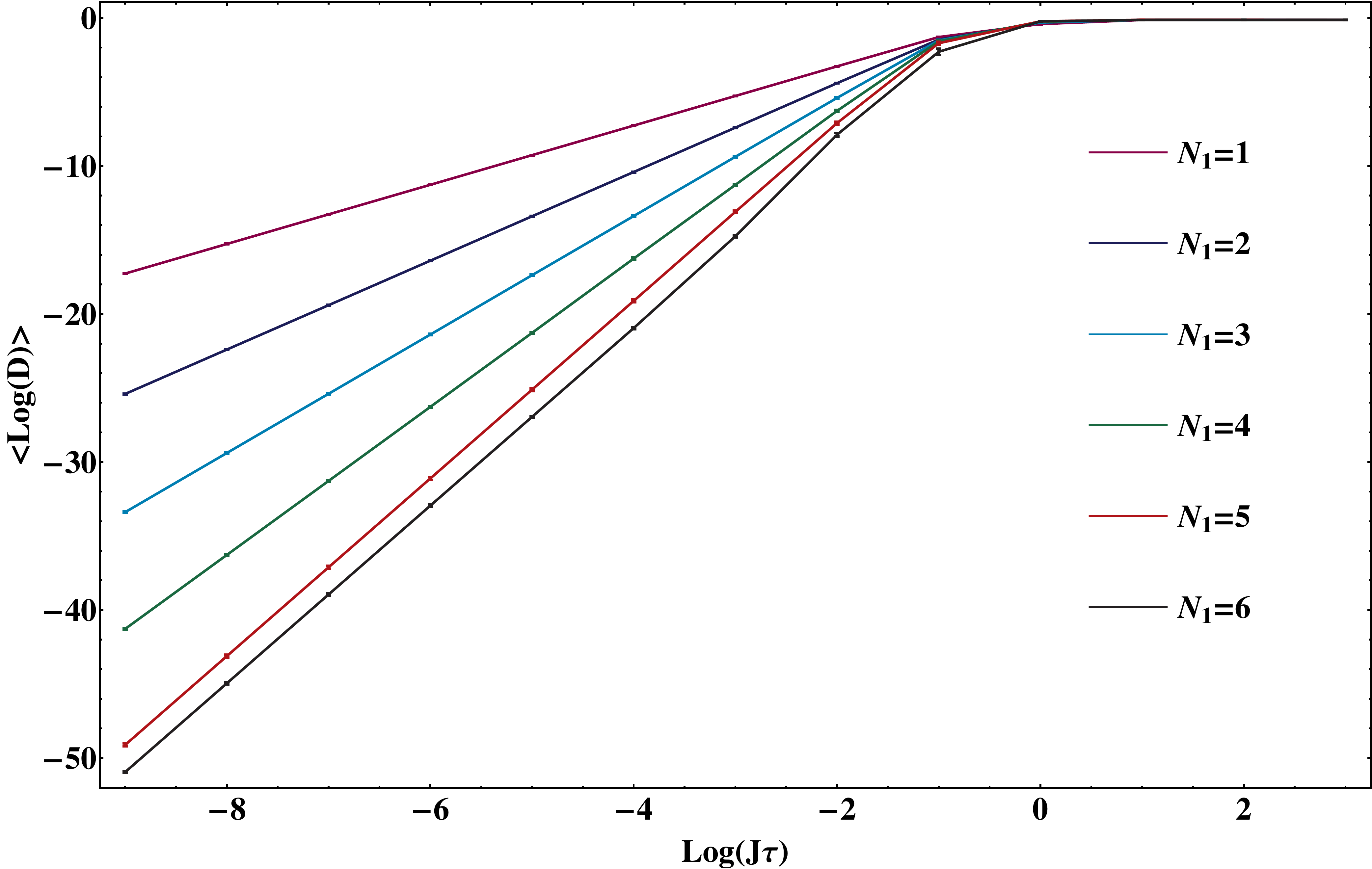}
\caption{(color online) Overall QDD distance measure after one cycle for $N_2=5$ and $%
N_1=1,2,\ldots,6 $ (left to right, top to bottom). See Fig.~\ref{fig:N2=3F} for additional details.}
\label{fig:N2=5dist}
\end{figure}

\begin{figure}[H]
\includegraphics[width=\columnwidth]{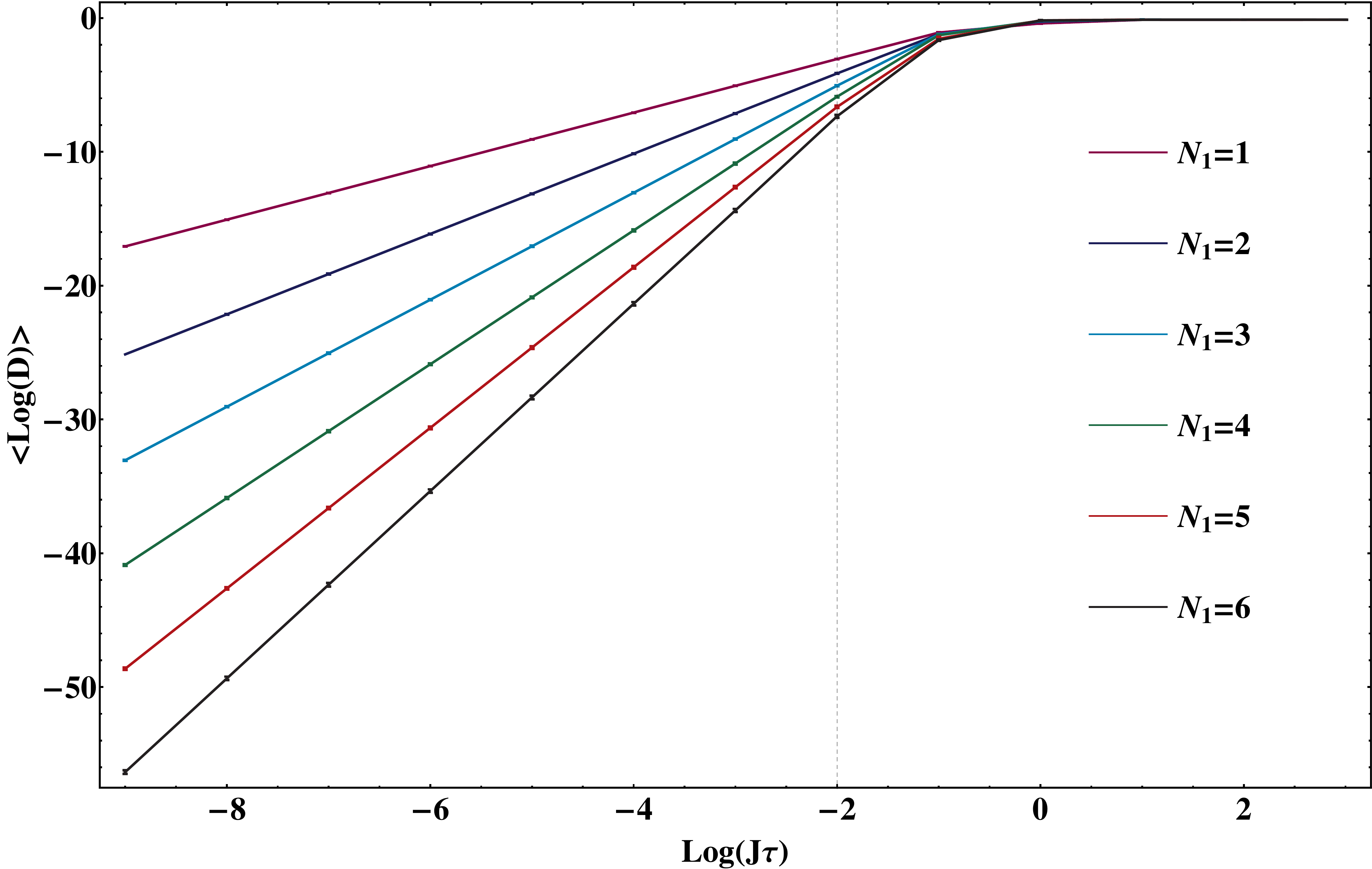}
\caption{(color online) Overall QDD distance measure after one cycle for $N_2=6$ and $%
N_1=1,2,\ldots,6 $ (left to right, top to bottom). See Fig.~\ref{fig:N2=4F} for additional details.}
\label{fig:N2=6dist}
\end{figure}

\begin{figure}[H]
\includegraphics[width=\columnwidth]{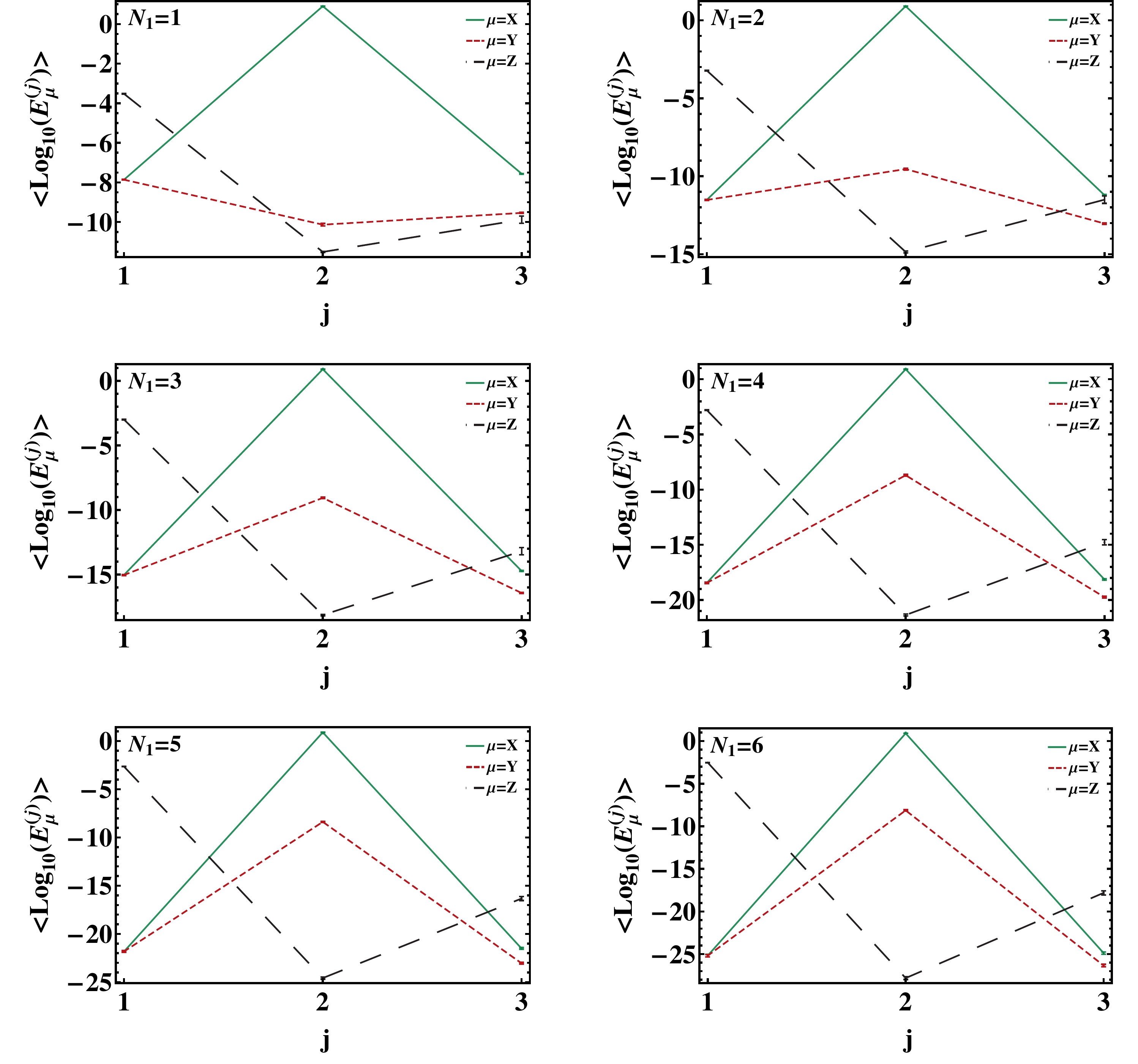}
\caption{(color online) Intermediate single-axis errors for $N_2=1$. See Fig.~\ref{fig:IntError3} for additional details}
\label{fig:IntError1}
\end{figure}

\begin{figure}[H]
\includegraphics[width=\columnwidth]{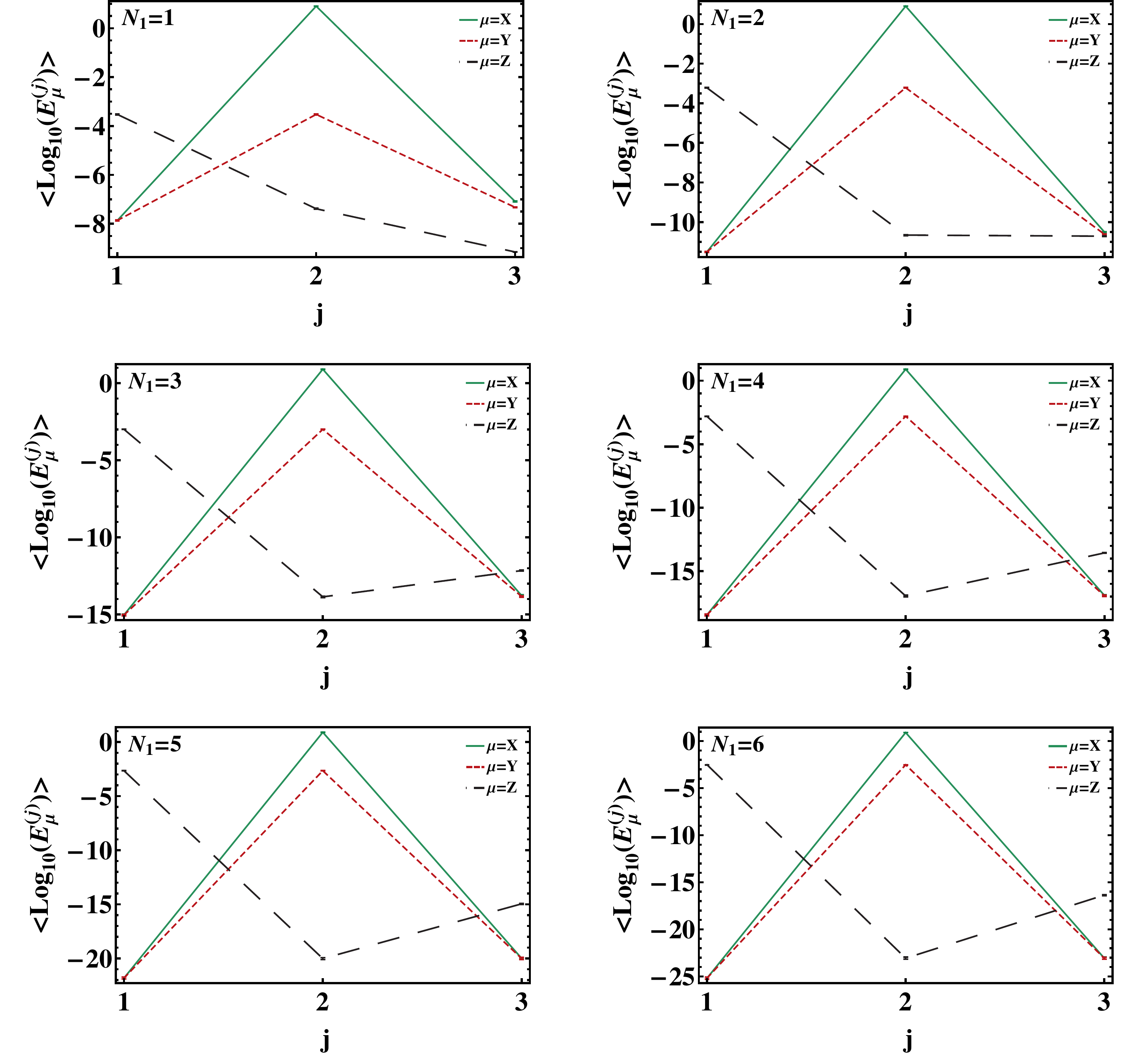}
\caption{(color online) Intermediate single-axis errors for $N_2=2$. See Fig.~\ref{fig:IntError4} for additional details}
\label{fig:IntError2}
\end{figure}

\begin{figure}[H]
\includegraphics[width=.95\columnwidth]{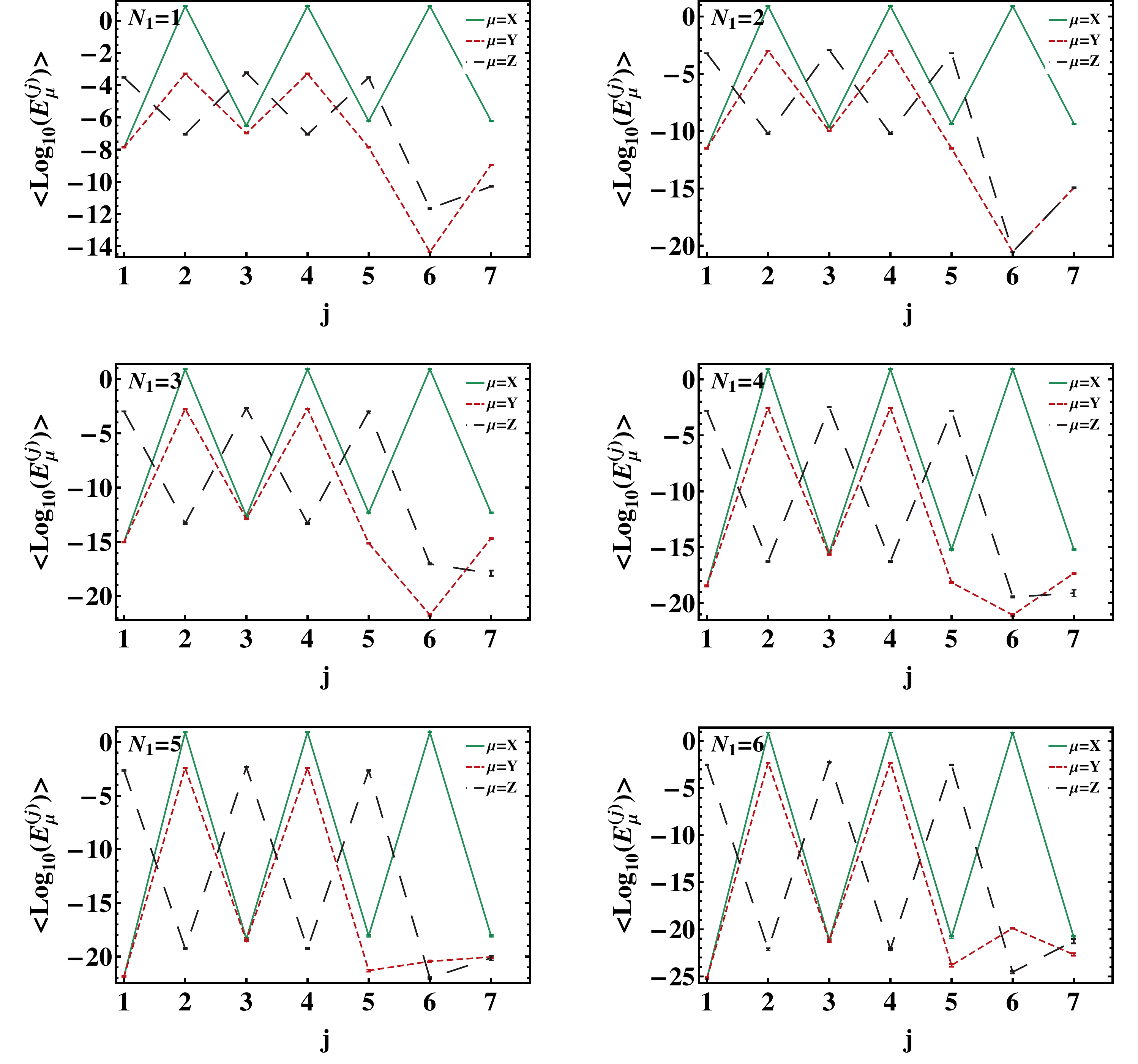}
\caption{(color online) Intermediate single-axis errors for $N_2=5$. See Fig.~\ref{fig:IntError3} for additional details}
\label{fig:IntError5}
\end{figure}

\begin{figure}[H]
\includegraphics[width=\columnwidth]{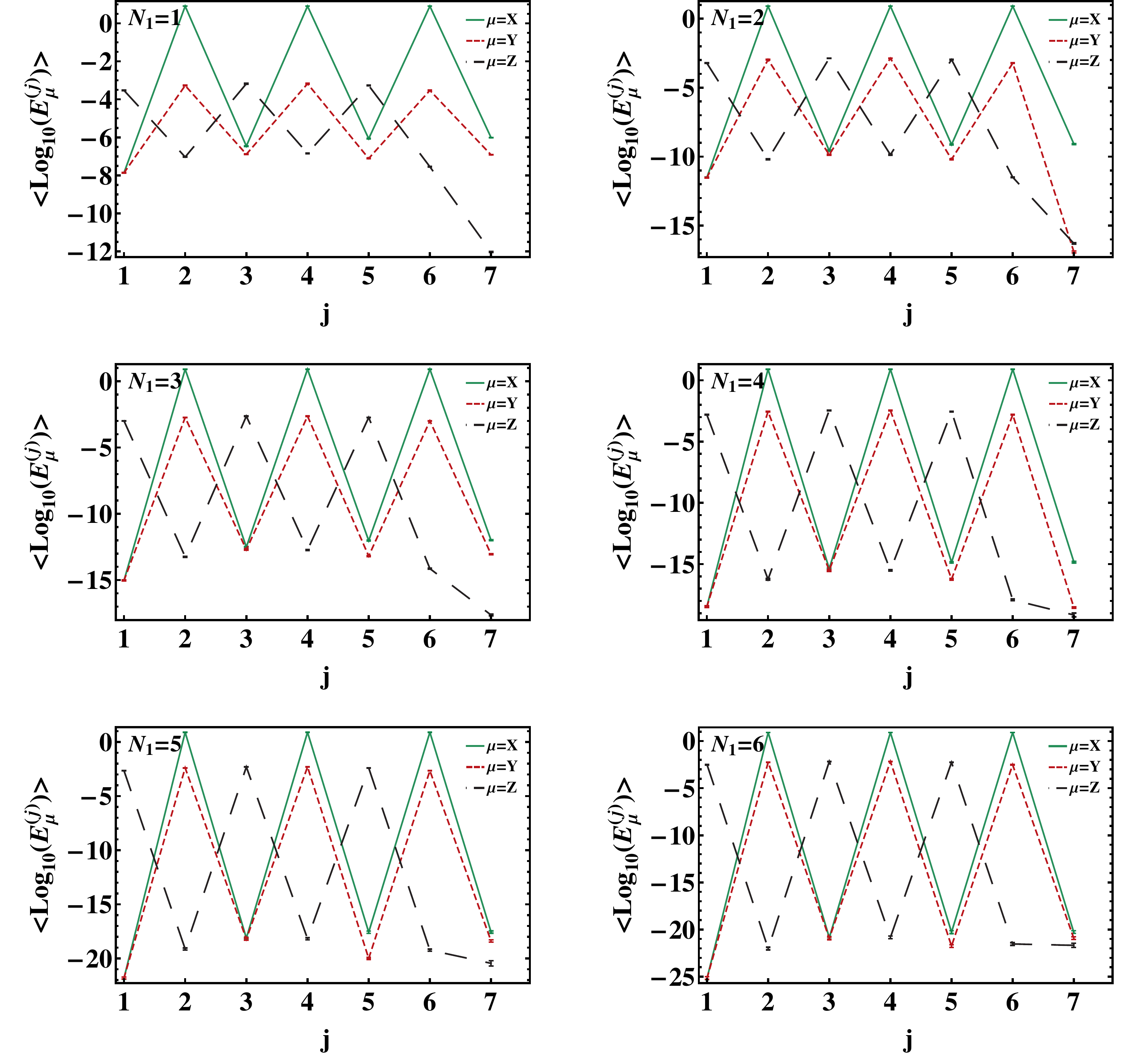}
\caption{(color online) Intermediate single-axis errors for $N_2=6$. See Fig.~\ref{fig:IntError4} for additional details}
\label{fig:IntError6}
\end{figure}

\end{document}